\documentclass[a4paper,hyper]{andp2012}
\usepackage[english]{babel}
\usepackage{amsmath,amssymb,amsfonts,amsthm,mathrsfs}
\usepackage{color}
\usepackage{url}

\let\normalcolor\relax

\newcommand{\be}{\begin{equation}}
\newcommand{\ee}{\end{equation}}
\newcommand{\ba}{\begin{eqnarray}}
\newcommand{\ea}{\end{eqnarray}}
\def\bs{\begin{subequations}}
\def\es{\end{subequations}}
\def\a{\alpha}
\def\b{\beta}
\def\g{\gamma}
\def\de{\delta}
\def\k{\kappa}
\def\e{\epsilon}
\def\vr{\varrho}
\newcommand{\V}{{\text{\tiny $V$}}}

\def\s{\sigma}

\def\dpl{\delta_{\rm Pl}}
\def\cR{{\cal R}}
\def\cH{\mathcal{H}}

\def\cN{{\cal N}}
\def\cV{{\cal V}}

\def\cP{{\cal P}}
\def\p{\partial}
\newcommand{\Eq}[1]{(\ref{#1})}
\def\lp{\ell_{\rm Pl}}
\def\rme{e}
\def\rmd{d}
\def\rmi{i}
\def\com{\color{magenta}}
\def\cob{\color{blue}}
\def\dpl{\delta_{\rm inv}}

\newcommand{\arX}[1]{\href{http://arxiv.org/abs/#1}{{\ttfamily\com arXiv:#1}}}
\newcommand{\doi}[2]{\href{http://dx.doi.org/#1}{\cob #2}}

\setcopyrightyear{2013}%
 \DOIprefix{10.1002}%
 \DOIsuffix{andp.201200227}%
 \Volume{525}%
 \Issue{5}%
 \Year{2013}%
 \pagespan{323}{338}
\Receiveddate{August 1, 2012}
\category{Review Article}

\title{Observational effects from quantum cosmology}
\author[G. Calcagni]{Gianluca Calcagni\inst{1,2,}\footnote{Corresponding author\quad E-mail:~\textsf{calcagni@iem.cfmac.csic.es}}}
\address[1]{Max Planck Institute for Gravitational Physics (Albert Einstein Institute),
Am M\"uhlenberg 1, D-14476 Golm, Germany}
\address[2]{Instituto de Estructura de la Materia, IEM-CSIC, Serrano 121, 28006 Madrid, Spain}

\begin{abstract}
The status of quantum cosmologies as testable models of the early universe is assessed in the context of inflation. While traditional Wheeler--DeWitt quantization is unable to produce sizable effects in the cosmic microwave background, the more recent loop quantum cosmology can generate potentially detectable departures from the standard cosmic spectrum. Thus, present observations constrain the parameter space of the model, which could be made falsifiable by near-future experiments.
\end{abstract}
\shortabstract

\begin{document}
\maketitle


\section{Introduction}

During the last years, quantum gravity has been receiving a great amount of attention from the community of theoreticians. The driving motivation, familiar to anyone who has tried his or her fortune at least once in this broad subject, is to realize a consistent, ultraviolet finite merging of general relativity with quantum mechanics. The programme can be carried out in various forms, from ambitious theories of everything (such as string theory) where all forces are unified to more minimalistic approaches aiming to quantize gravity alone. In the latter category there fall loop quantum gravity (LQG), asymptotic safety, spin-foams, causal dynamical triangulations and many others \cite{Ori09}.

A problem endemic to most of these scenarios is their difficulty in making contact with observations. This stems from the highly technical nature of the theoretical frameworks, where the notions of conventional geometry and matter, continuum spacetime, general covariance and physical observables are typically deformed, modified, or disappear altogether. The lack of experimental feedback makes it quite difficult to discriminate among different models and, chiefly, to characterize them as \emph{falsifiable}.

It is natural to turn to cosmology in an attempt to bridge this gap and advance our knowledge \cite{Kie07,Boj11}. The early Universe is an ideal laboratory where extreme regimes of high energy and high curvature are realized. Under such conditions, it is expected that quantum gravitational effects become sizable. Also, the symmetry reduction entailed in cosmological settings decimates the degrees of freedom of background-independent theories and allows one to simplify the latter to a technically manageable level. The resulting models retain some (or most) of the main features of the full theory and can be better manipulated to extract observables.

Canonical quantum gravity is a popular example of this mechanism. The present review focusses on two of its incarnations, namely, the traditional Wheeler--DeWitt (WDW) model (e.g., \cite{Wil95,CQC}) and the more recent loop quantization \cite{Boj08}. The most ancient phase about which we have gathered experimental data is inflation, a period of accelerated expansion of the universe which left a relic in the cosmic microwave background (CMB) radiation. A study of the inflationary perturbations and the associated spectra allows us to track down quantum corrections and confront them with the observed CMB power spectrum. Although the outcome of this procedure is a constraint on the free parameters of the models rather than an actual prediction, time seems ripe for the very next generation of experiments to exclude notable portions of parameter space. As a minimal present-day achievement, we can at least state that quantum cosmology models are compatible with observations.

The stark contrast between the type of quantum corrections arising in these scenarios highlights how sensitive the physics is of the quantization scheme and variables. The typical energy scale during inflation is estimated to be about the grand-unification scale, $H\sim 10^{15}\,\mbox{GeV}$, corresponding to an energy density $\rho_{\rm infl}\sim H^2/\lp^2\sim 10^{68}\,\mbox{GeV}^4$. Here $H:=\dot a/a$ is the Hubble parameter, $a$ is the scale factor of the universe and a dot denotes differentiation with respect to synchronous time. In contrast, classical gravity is believed to break down at distances shorter than the Planck length $\lp=\sqrt{G\hbar}$, i.e., at energies above $10^{19}\,\mbox{GeV}$. The ratio between the inflationary and Planck energy density is very small,
\be\label{wdwqc} 
\frac{\rho_{\rm infl}}{\rho_{\rm Pl}}\sim (\lp H)^2 \sim 10^{-8}\,,
\ee
and quantum corrections are expected to be of the same order of magnitude or lower. Thus, quantum-gravity effects would be, in fact, well below any reasonable experimental sensitivity threshold, at least as far as inflation is concerned. WDW quantum cosmology realizes precisely this type of corrections and endorses the above naive argument. 

On the other hand, the polymeric quantization of loop quantum cosmology (LQC) \cite{CQC,lqcr1,lqcr2} generates corrections which are not of the form \Eq{wdwqc}. To get a rough idea of how these corrections arise, one begins by observing that geometry operators representing areas and volumes acquire a discrete spectrum in this context. This is because states of loop quantum gravity, spin networks, are graphs whose edges $e$ are labeled by quantum numbers $j_e$. An area intersected by some of these edges is determined by these quantum numbers, giving the spectrum ${\cal A}=\gamma \lp^2 \sum_e\sqrt{j_e(j_e+1)}$, where $\g\lesssim O(1)$ is the Barbero--Immirzi parameter. One single edge defines an ``elementary plaquette'' of area $\propto \lp^2 \sqrt{j_e(j_e+1)}$; the latter features the Planck area but its actual value depends on the spin quantum number. Since calculations on realistic graphs are very hard in the full theory, it is convenient to focus one's attention on a simplified phenomenological setting. In particular, a homogeneous quantum inflationary universe with small inhomogeneous perturbations may be represented by a quantum semi-classical state $\Psi$ characterized by a length scale $L$. This scale is thought of as encoding the discreteness of the geometry. Any region of volume $\cV=a^3\cV_0$ (arbitrary, if spatial slices are non-compact) can be decomposes into discrete patches of size $\sim L^3$. The inflationary scale is thus replaced by an effective quantum-gravity scale 
\be\label{rhoqg}
\rho_\textsc{qg}= \frac{3}{8\pi G L^2}\,.
\ee

In general, inverse powers of $L$ cannot be quantized to a densely defined operator because the spectrum of the volume contains 0. Inverse volumes appear in the Hamiltonian constraint (of both gravity and matter, as in kinetic matter terms) and hence in the dynamics, and are an unavoidable consequence of spatial discreteness in loop quantum gravity. This requires to reexpress their classical expressions via Poisson brackets, which in turn feature derivatives by $L$. Quantum discreteness then replaces classical continuous derivatives by finite-difference quotients. For example, the expression $(2\sqrt{L})^{-1}=\p\sqrt{L}/\p L$ would become $(\sqrt{L+\lp}-\sqrt{L-\lp})/(2\lp)$, strongly differing from $(2\sqrt{L})^{-1}$ 
when $L$ is as small as the Planck length, $L\sim\lp$. For larger $L$, corrections
are perturbative and of the order $\lp/L$, so in general the type of inverse-volume quantum corrections are expressed by the ratio
\be\label{lqcqc}
\frac{\rho_\textsc{qg}}{\rho_{\rm Pl}}\sim \left(\frac{\lp}{L}\right)^2 \lesssim 1\,.
\ee
In practice, the actual size of LQC effects will lie well below the over-optimistic upper bound \Eq{lqcqc}, but above the naive estimate \Eq{wdwqc}. It is known that the non-local nature of loop quantum gravity effects prevents the formation of singularities one would typically find classically \cite{Boj08,Bo06b,Ash08}. This can be shown both at the kinematical level (via the spectra of inverse area and volume operators) and at the exact and effective dynamical level (by looking, respectively, at the state-space spanned by the Hamiltonian constraint acting on volume eigenstates and at the effective dynamics on semi-classical states). The physical interpretation of inverse-volume corrections stems exactly from the same mechanism: classically divergent quantities such as inverse powers of volumes remain finite due to intrinsically quantum effects. Loosely speaking, quanta of geometry cannot be compressed too densely and they determine the onset of a repulsive force at Planck scale \cite{Ash08}, which then determine the various corrections to the dynamics.

After introducing the theoretical frameworks in sections \ref{th1} and \ref{th2}, CMB observations will be used to pin down these effects (sections \ref{ex1} and \ref{ex2}). For the Wheeler--DeWitt model, we shall do so in considerably more detail than can be found in the present literature; section \ref{ex1} contains original material. Holonomy corrections in LQC are briefly discussed in section \ref{th3} The scantly touched topic of non-Gaussianity in quantum cosmology will be also discussed (section \ref{nonga}). In the following, $\hbar=1=c$.

Before starting, we stress once again the scope of the present review. Although there are many ``minimalistic'' theories of quantum gravity on the market, at present it is still difficult to do some cosmology with them. Among the scenarios allowing for some phenomenology are asymptotic safety \cite{ReS} and causal dynamical triangulations \cite{Gor13}. These models do admit a cosmological limit, but either inflationary observables have not been computed yet or there is no unique determination of an effective inflationary gravitational action. Here, on the other hand, we are interested in pitching models based upon canonical quantization (which conventionally go under the umbrella term ``quantum cosmology'') against observations. We will leave out string cosmology from the discussion \cite{BM,MuW}, which is based on altogether different techniques. The reader can find the details of various and often interconnected settings in the dedicated literature, such as KKLT and moduli inflation \cite{Kal07,CQ}, cosmic strings networks \cite{CPV,Avg11}, brane and DBI inflation \cite{BM,MuW}, string gas cosmology \cite{Bra11,Bra12}, braneworld cosmology \cite{MaK}, ekpyrotic universe \cite{Leh08,Leh11}, non-local cosmology, and others.


\section{Wheeler--DeWitt cosmology and observations}\label{wdw}


\subsection{The model}\label{th1}

\subsubsection{Homogeneous background}

In canonical formalism, symmetry and dynamics are encoded in a set of constraint equations valid on dynamical trajectories. For gravity and matter, the total Dirac Hamiltonian \cite{HTe} obtained after imposing second-class constraints and skimming out Lagrange multipliers is
\be\label{hd}
H_{\rm D}=\int \rmd^3{\bf x}\,\left(N^\a\cH_\a+N\cH\right)\,,
\ee
where $N^\a$ ($\a=1,2,3$) is the shift vector, $N$ is the lapse function, $\cH_\a$ is the super-momentum constraint and $\cH$ is the super-Hamiltonian constraint (often the prefix ``super'' is omitted). The super-momentum, corresponding to the 0$\a$ components of Einstein's equations, encodes invariance under spacetime diffeomorphisms within the three-dimensional spatial surfaces on which one integrates. The super-Hamiltonian (the 00 component of Einstein's equations) both encodes invariance under time reparametrizations and generates the dynamics (time evolution) of the system. Symmetry and dynamics are thus entangled. Canonical quantization follows by promoting the first-class constraints $\cH_\a$ and $\cH$ to operators acting on a Hilbert space of wave-functionals $\Psi$. Quantum dynamics is then fully specified by the equations $\hat\cH_\a\Psi=0$ and the Wheeler--DeWitt equation
\be\label{wdwe}
\hat\cH\Psi=0\,.
\ee

In a fully background-independent theory, both $\hat\cH_\a$ and $\hat\cH$ are written in terms of the canonical variables associated with the fundamental degrees of freedom (metric and matter) of the system. These expressions are non-linear and, in practice, it is extremely difficult to solve the constraint equations and construct the physical Hilbert space. Symmetry reduction (at the classical level) to the flat, homogeneous and isotropic Friedmann--Lema\^itre--Robertson--Walker (FLRW) metric $g_{\mu\nu}=(-1,a^2(t),a^2(t),$ $a^2(t))$ greatly simplifies the problem. The momentum constraint is composed only of spatial derivatives and it vanishes identically. After integrating over the spatial volume (formally divergent but regularizable), the super-Hamiltonian in the presence of a matter scalar field $\phi$ with potential $V$ reads
\be\label{cch}
\cH=\frac{1}{2a^3}\left[-\frac{a^2 p_{(a)}^2}{6\k^2}+\Pi_\phi^2\right]+a^3\left[V(\phi)-\frac{3}{\k^2}\frac{\textsc{k}}{a^2}\right]\,,
\ee
where $p_{(a)}=-6a\dot a/N$ and $\Pi_\phi =a^3\dot\phi/N$ are the momenta conjugate to $a$ and $\phi$, respecively, $\k^2=8\pi G$, and $\textsc{k}=0,\pm1$ is the curvature of spatial slices. The constraint $\cH=0$ is nothing but the first Friedmann equation
\be\label{1fe}
H^2=\frac{\k^2}{3}\left[\frac{\dot\phi^2}2+V(\phi)\right]-\frac{\textsc{k}}{a^2}\,.
\ee
The other classical equation of motion is that for the scalar field,
\be\label{kge}
\ddot\phi+3H\dot\phi+V_{,\phi}(\phi)=0\,.
\ee

Quantizing expression \Eq{cch} and promoting $a$ and $\phi$ to multiplicative operators and the momenta to derivative operators $\hat p_{(a)}:=-\rmi\p_a$ and $\hat\Pi_\phi:=-\rmi\p_\phi$, one obtains $\hat\cH\Psi[\cN,\phi]=0$, where
\be\label{miniHq2}
\hat\cH=\frac{\rme^{-3\cN}}{2}\left[\frac{\k^2}{6}\frac{\p^2}{\p\cN^2}-\frac{\p^2}{\p\phi^2}+2\rme^{6\cN}V(\phi)-\frac{6\textsc{k}}{\k^2}\rme^{4\cN}\right]
\ee
and $\cN=\ln a$ is the number of e-foldings. This equation may not necessarily be regarded as fundamental.\footnote{Apart from the issue of symmetry reduction, the actual quantization of the putative full theory can lead to an altogether different expression; LQG is an example. Also, canonical quantum gravity may be embedded in a more general field-theory approach such as group field theory \cite{Ori09}. The wave-function $\Psi$ is promoted to a field and the Wheeler--DeWitt equation \Eq{wdwe} receives non-linear corrections. Linearizing, one gets an effective Hamiltonian which can be considered also at the level of mini-superspace \cite{CGO2}.} However, it gives the correct result in the semi-classical limit, and one can assume it as an effective description of the quantum universe in this regime.

During inflation, the scalar field varies very slowly and its kinetic term is negligible with respect to the potential (slow-roll regime); at the quantum level, it corresponds to dropping the $\p_\phi^2$ term in Eq.\ \Eq{miniHq2}. Assuming a quadratic potential $V(\phi)=\tfrac12 m^2\phi^2$, from the Friedmann equation \Eq{1fe} it follows that
\be\label{sra}
\frac{6H^2}{\k^2}\approx m^2\phi^2\,.
\ee
Thus, the energy scale of inflation sets what shall later play the role of quantum correction.

\subsubsection{Perturbations}

When inhomogeneities are switched on, the FLRW mini-superspace framework breaks down and one should consider the full Dirac Hamiltonian \Eq{hd}. Since the super-momentum and super-Hamiltonian constraints are non-linear in the canonical variables,
the problem quickly becomes intractable unless one resorts to some approximations. Inflationary inhomogeneous fluctuations are very small, so linear perturbation theory is sufficient to obtain the spectra. The matter scalar is decomposed into a homogeneous background (representing the vacuum expectation value of the field) and a fluctuation, $\phi(t,{\bf x})=\phi(t)+\de\phi(t,{\bf x})$. In this section we ignore the metric backreaction $\de g_{\mu\nu}$, in which case the scalar is regarded as a ``test'' field. In the standard cosmological model, backreaction does not affect the power spectrum at lowest order in perturbation theory and in the slow-roll truncation. This suffices for our purposes also in WDW quantum cosmology. (However, we shall include backreaction in the LQC case.) The scalar perturbation is decomposed into Fourier modes,
\be 
\de\phi(t,{\bf x})=\sum_{{\bf k}} \de\phi_k(t)\,\rme^{\rmi {\bf k}\cdot {\bf x}},
\ee
where we assumed spatial slices to be compact ($\textsc{k}=1$) and the Fourier mode depends only on the modulus $k=|{\bf k}|$. 
 Replacing $\phi(t)$ with $\phi(t,{\bf x})$ in the WDW equation \Eq{miniHq2}, the mini-superspace is extended to include also the infinity of modes $\de\phi_k$. The wave-function $\Psi[\cN,\phi,\{\de\phi_k\}_k]$ can be actually factorized as a background part times the rest, $\Psi[\cN,\phi,\{\de\phi_k\}_k]=\Psi_0[\cN,\phi]\prod_{k>0}\Psi_k[\cN,\phi,\de\phi_k]$. In doing so, one drops self-interaction terms which are consistently negligible in first-order perturbation theory. Eventually, one obtains \cite{HaH,Kie87}
\ba
&&\frac{\rme^{-3\cN}}{2}\left[\frac{\k^2}{6}\frac{\p^2}{\p\cN^2}-\frac{\p^2}{\p\de\phi_k^2}
+\rme^{6\cN}\frac{6H^2}{\k^2}\right.\nonumber\\
&&\qquad+\left.\vphantom{\frac{\k^2}{6}}\left(\rme^{6\cN}m^2+\rme^{4\cN}k^2\right)\de\phi_k^2\right]\psi_k[\cN,\de\phi_k]\approx 0\,,\label{bo}
\ea
where $\psi_k[\cN,\de\phi_k]=\Psi_0[\cN,\phi]\Psi_k[\cN,\phi,\de\phi_k]$ and the $\phi$ dependence is omitted because we used the slow-roll approximation \Eq{sra} to express the background potential in terms of the Hubble parameter.

Noting that $\cN$ and $\de\phi_k$ correspond, respectively, to slow- and fast-evolving variables, at this point one can make a Born--Oppenheimer approximation on the solution \cite{Kie07,KiS}. The latter is written as 
\be\label{ans}
\psi_k[\cN,\de\phi_k]=\exp[\rmi S(\cN,\de\phi_k)]
\ee
and the functional $S$ is expanded in $m_{\rm Pl}^2=3/(2\pi\lp^2)$ $=12/\k^2$: $S=m_{\rm Pl}^2 S_0+ S_1+m_{\rm Pl}^{-2}S_2+\dots$. Plugging the \emph{Ansatz} \Eq{ans} into Eq.\ \Eq{bo} and expanding, the $O(m_{\rm Pl}^4)$ and $O(m_{\rm Pl}^2)$ terms imply $S_0=\pm\rme^{3\cN}H/6$, while at the next two orders one finds two equations for the wave-functions
\ba
\psi_k^{(0)}[\cN,\de\phi_k]&:=& A(\cN)\,\rme^{\rmi S_1(\cN,\de\phi_k)}\,,\\
\psi_k^{(1)}[\cN,\de\phi_k]&:=& B(\cN)\psi_k^{(0)}[\cN,\de\phi_k]\,\rme^{\rmi m_{\rm Pl}^{-2} S_2(\cN,\de\phi_k)},
\ea
where $A$ and $B$ are chosen to match the amplitudes in the WKB approximation.

\subsubsection{Observables}

The wave-functions $\psi_k^{(0)}$ and $\psi_k^{(1)}$ have been computed semi-analytically in \cite{KK1,KK2}, to which we refer the reader for details. From the explicit solutions, one can calculate the two-point correlation function
\be\label{Pn} 
P_\phi^{(n)}(k):=\langle\psi_k^{(n)}||\de\phi_k|^2|\psi_k^{(n)}\rangle
\ee
of the scalar perturbation order by order. This quantity is directly related to the imprint of inhomogeneous fluctuations in the cosmic microwave background. However, only perturbations which left the comoving Hubble horizon $(aH)^{-1}=:k_*^{-1}$ and later reentered it can be observed in the sky. Therefore, the actual cosmological observable is Eq.\ \Eq{Pn} in the long wave-length limit $k\ll k_*$, then evaluated at $k=k_*$. This is the $n$-th order power spectrum
\be 
\cP_{\rm s}^{(n)}(k):=\frac{k^3}{2\pi^2}P_\phi^{(n)}(k\ll k_*)\big|_{k=k_*}\,.
\ee
The lowest-order result coincides with the standard one,
\be\label{cps}
\cP_{\rm s}^{(0)}=\frac{\k^2}{2}\frac{1}{\e}\left(\frac{H}{2\pi}\right)^2\,,
\ee
where 
\be\label{1srp}
\e:=-\frac{\dot H}{H^2}=\frac{\k^2}{2}\frac{\dot\phi^2}{H^2}
\ee
is the first slow-roll parameter. Since both $H$ and $\e$ are approximately constant during inflation, the spectrum (which we sometimes call ``classical'' because of the absence of quantum-gravity corrections) is almost scale invariant.

The next-to-lowest-order expression is the standard one times a quantum correction \cite{KK1}:
\be\label{cP10wdw} 
\cP_{\rm s}(k)\approx \cP_{\rm s}^{(1)}(k)=\cP_{\rm s}^{(0)}(k) C_k^2\,,
\ee
where
\bs\ba  
C_k^2 &\approx& \left(1-\frac{43.56}{k^3}\frac{H^2}{m_{\rm Pl}^2}\right)^{-3}\left(1-\frac{189.18}{k^3}\frac{H^2}{m_{\rm Pl}^2}\right)^2\label{ck}\\
    &=& 1-\frac{247.68}{k^3}\frac{H^2}{m_{\rm Pl}^2} +\frac{1}{k^6}O\left(\frac{H^4}{m_{\rm Pl}^4}\right)\nonumber\\
    &\approx& 1-\de_\textsc{wdw}(k)+O(\de_\textsc{wdw}^2)\,,
\ea\es
and we dubbed the leading Wheeler--DeWitt quantum correction
\be\label{dewdw}
\de_\textsc{wdw}(k):=\frac{10^3}{k^3}(\lp H)^2\,.
\ee
$C_k\to 1$ in the small-scale limit ($k\to\infty$), while at large scales ($k\ll k_*$) the quantum-corrected power spectrum acquires a mild scale dependence which makes the signal \emph{suppressed} with respect to the standard result. A similar suppression of the spectrum happens also in other models where geometry is quantized, such as non-commutative and string inflation \cite{TMB,PTZ,Cal04,CT1}. At first, it might seem counter-intuitive that quantum gravity affects large scales more than small scales. However, large-scale perturbations left the horizon before (and hence reentered after) smaller-scale fluctuations, and they were longer exposed to high-energy and high-curvature effects. The approximation scheme used to derive Eq.\ \Eq{ck} breaks down in the limit $C_k\to 0$ and the critical $k$ at which that happens should not be taken as a physical threshold.

From the power spectrum, one can compute the scalar spectral index
\be\label{ns}
n_{\rm s}-1:= \frac{\rmd\ln \cP_{\rm s}}{\rmd\ln k}\,,
\ee
which generalizes the definition of an exactly power-law-type spectrum $\cP_{\rm s}\sim k^{n_{\rm s}-1}$. To calculate this, we notice that (from $aH=k$ at horizon crossing) $\rmd/\rmd\ln k\approx \rmd/(H\rmd t)$ and we recall the background relations, stemming from the equations of motion,
\be
\dot\e   = 2H\e(\e-\eta)\,,\qquad \dot\eta =  H(\e\eta-\xi^2)\,,
\ee
where
\be
\eta  := -\frac{\ddot\phi}{H\dot\phi}\,,\qquad \xi^2 := \frac{1}{H^2} \left(\frac{\ddot{\phi}}{\dot{\phi}}\right)^.= \frac{\dddot{\phi}}{H^2\dot{\phi}}- \eta^2\,,
\ee
are the second and third slow-roll parameter, respectively. Since $H\approx {\rm const}$, one gets
\be\label{eigenwdw} 
\frac{\rmd\de_\textsc{wdw}}{\rmd\ln k}\approx -3\de_\textsc{wdw}
\ee
and
\be\label{nswdw} 
n_{\rm s}-1 \approx 2\eta-4\e+ 3\de_\textsc{wdw}\,,
\ee
where we have dropped higher-order terms in the combined $\de_\textsc{wdw}$/slow-roll expansion. Positivity of the quantum correction in Eq.\ \Eq{nswdw} ensures suppression of power at low wavenumbers.

The next slow-roll observable is the running of the spectral index:
\be\label{as}
\a_{\rm s}:= \frac{\rmd n_{\rm s}}{\rmd\ln k}\,.
\ee
Combined with Eqs.\ \Eq{eigenwdw} and \Eq{nswdw}, it leads to
\be\label{aswdw}
\a_{\rm s}\approx 2\left(5\e\eta-4\e^2-\xi^2\right)-9\de_\textsc{wdw}\,.
\ee

The scalar power spectrum expanded to all orders in the perturbation wavenumber 
about a pivot scale $k_0$ is
\ba
\ln {\cal P}_{\rm s} (k) &=&\ln {\cal P}_{\rm s} (k_0)+[n_{\rm s}(k_0)-1] x
+\frac{\alpha_{\rm s}(k_0)}{2}x^2\nonumber\\
&&+\sum_{m=3}^{\infty} \frac{\alpha_{\rm s}^{(m)}(k_0)}{m!}x^m\,,
\label{Ps2}
\ea
where $x := \ln (k/k_0)$. As the order of the observables
\be \label{use}
\alpha_{\rm s}^{(m)}:= \frac{\rmd^{m-2} \a_{\rm s}}{(\rmd\ln k)^{m-2}}\approx O(\e^m)-(-3)^m\de_\textsc{wdw}
\ee
increases, the classical part becomes smaller and smaller but the leading-order quantum correction survives. At some order $m$, the quantum correction will dominate over the standard part. Taking \Eq{use} into account, Eq.\ \Eq{Ps2} can be recast as
\be 
\ln {\cal P}_{\rm s} (k) \approx\ln {\cal P}_{\rm s}^{(0)} (k)+\de_\textsc{wdw}(k_0)\left[1-\left(\frac{k_0}{k}\right)^3\right]\,.\label{Ps3}
\ee


\subsection{Experimental bounds}\label{ex1}

Equation \Eq{wdwqc} is written in units where $k$ is dimensionless. In fact, one should make the replacement $k\to k/k_{\rm min}$, where $k_{\rm min}\sim 1.4\times 10^{-4}\,\mbox{Mpc}^{-1}$ is the largest observable scale. Here we used the fact that comoving wavenumbers and multipoles are approximately related by $k\approx \ell/\tau_0$, where $\tau_0\approx 14.4\,\mbox{Gpc}$ is the comoving particle horizon today, and that the lowest early-universe contribution to the CMB spectrum is the quadrupole $\ell=2$. One can reexpress $\de_\textsc{wdw}$ in terms of spherical multipoles, and $k/k_{\rm min}=\ell/\ell_{\rm min}=\ell/2$. A more generous estimate for the quantum correction will stem by replacing $k_{\rm min}$ by the pivot scale $k_0\gg k_{\rm min}$, which we adopt from now on: $k\to k/k_0$.

The WMAP7 mean for the scalar amplitude in the absence of tensor signal is $\cP_{\rm s}(k_0) = (2.43\pm 0.11)\times 10^{-9}$ at $k_0=0.002\,\mbox{Mpc}^{-1}$ and 68\% confidence level (C.L.) \cite{Lar10}, where the pivot scale $k_0$ corresponds to a CMB multipole $\ell_0\approx 29$. Equation \Eq{cps} and the inflationary condition $\e<1$ yield the upper bound
\be\label{infbo}
(\lp H) < 9 \times 10^{-5}\,,
\ee
as anticipated in Eq.\ \Eq{wdwqc}. The bound can be recast for the Hubble parameter alone, $H< 3.2\times 10^{15}\,\mbox{GeV}$, or, via the classical equation of motion \Eq{1fe} in the slow-roll approximation, for the inflaton potential, $V^{1/4} < 6.8\times 10^{16}\,\mbox{GeV}$. In particular, the WDW quantum correction is constrained to be
\be\label{dwdw}
\de_\textsc{wdw}(k_0)< 7.9\times 10^{-6}\,.
\ee
With $k_{\rm min}$ instead of $k_0$ the quantum correction is further suppressed, $\de_\textsc{wdw}(k_0)< 2.6\times 10^{-9}$.

Even taking the upper bound $\lp H = 10^{-4}$, $\de_\textsc{wdw}= 10^{-5}$, quantum corrections are too small to be detected. Their dependence on the inflationary energy scale is crucial for this result. Another reason is that at large scales cosmic variance is the leading source of error. The latter is a manifestation of the failure of the ergodic theorem for the discrete CMB multipole  spectrum. For the power spectrum ${\cal P}_{\rm s}(\ell)$, cosmic variance is given by \cite{var1,var2}
\be
{\rm Var}_{{\cal P}_{\rm s}}(\ell)=\frac{2}{2\ell+1}\,{\cal P}_{\rm s}^2(\ell)\,.
\ee
Quantum-gravity corrections should be compared with the error bars due to 
cosmic variance with respect to the classical spectrum ${\cal P}_{\rm s}^{(0)}(\ell)$. The latter, Eq.\ \Eq{Ps2}, is determined up to the 
normalization ${\cal P}_{\rm s} (\ell_0)$, so that the region in the 
$(\ell,{\cal P}_{\rm s} (\ell)/{\cal P}_{\rm s} (\ell_0))$ plane 
affected by cosmic variance is roughly delimited by the two curves
\be
\frac{{\cal P}_{\rm s}^{(0)} (\ell)\pm \sqrt{{\rm Var}_{{\cal P}_{\rm s}^{(0)}}(\ell)}}
{{\cal P}_{\rm s}^{(0)} (\ell_0)}=\left(1\pm\sqrt{\frac{2}
{2\ell+1}}\right)\frac{{\cal P}_{\rm s}^{(0)} (\ell)}
{{\cal P}_{\rm s}^{(0)} (\ell_0)}\,,\label{cosv}
\ee
where we take the classical spectrum as reference. The WDW-corrected spectrum is given by Eq.\ \Eq{Ps3}. In the absence of tensor modes and running, the WMAP+BAO+$H_0$ dataset (combination of WMAP7 data and observations of baryon acoustic oscillations and the Hubble expansion) yields a scalar spectral index $n_{\rm s}(k_0)=0.963 \pm 0.012$ at $k_0=0.002\,\mbox{Mpc}^{-1}$ and 68\% C.L.\ \cite{Kom10}. This number can change depending on the priors, but not much. Classically, this corresponds to slow-roll parameters at most of order $\e\sim O(10^{-2})$. Therefore, the standard spectrum in Eq.\ \Eq{Ps3} can be approximated by $\ln {\cal P}_{\rm s}^{(0)} (k) \approx\ln {\cal P}_{\rm s} (k_0)+[n_{\rm s}(k_0)-1] x+\tfrac12\alpha_{\rm s}(k_0)x^2$. To plot the WDW spectrum, we only need to plug in values for the scalar index and its running. First, we recast the observables in terms of a set of slow-roll parameters dependent on the field potential (e.g., \cite{CQC}):
\be
\e_\V:=\frac{1}{2\k^2}\left(\frac{V_{,\phi}}{V}\right)^2,
\quad 
\eta_\V:=\frac{1}{\k^2}\frac{V_{,\phi\phi}}{V}\,,
\quad
\xi_\V^2 := \frac{V_{,\phi}V_{,\phi\phi\phi}}{\kappa^4 V^2}\,.\label{vsr}
\ee
The scalar index \Eq{nswdw} and its running \Eq{aswdw} become
\ba
n_{\rm s}-1 &=& -6\e_\V+2\eta_\V+ 3\de_\textsc{wdw}\,,\label{ns2} \\
\alpha_{\rm s} &=& -24 \e_\V^2+16\e_\V\eta_\V
-2\xi_\V^2-9\de_\textsc{wdw}\,.\label{alphas}
\ea
For a quadratic potential $V(\phi)\propto\phi^2$, 
\be
\e_\V=\frac{2}{\kappa^2 \phi^2}\,,
\qquad
\eta_\V=\e_\V\,,
\qquad
\xi_\V^2=0\,.
\ee
This allows one to reduce the slow-roll parameters to just one. A realistic theoretical value for $\e_\V$ at the pivot scale is $\e_\V(k_0)=0.009$. 

\begin{figure}
\includegraphics[width=\columnwidth]{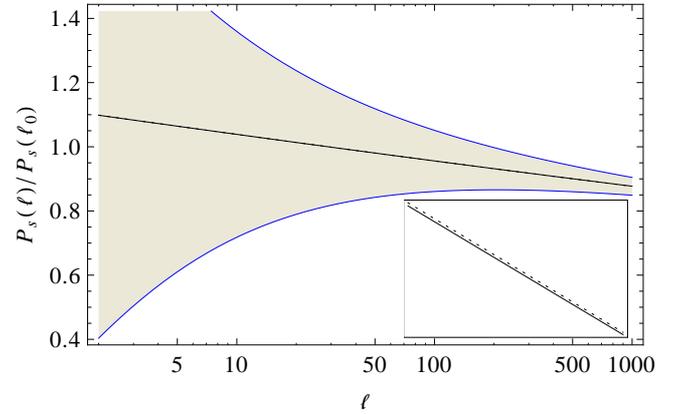}
 \caption{\label{fig1}
Log-linear plot of the Wheeler--DeWitt primordial scalar spectrum ${\cal P}_{\rm s}(\ell)$ for a quadratic inflaton potential, with $\epsilon_{\V}(k_0)=0.009$ and for the pivot wavenumber $k_0=0.002$\,Mpc$^{-1}$, corresponding to $\ell_0= 29$. The shaded region, delimited by the two curves \Eq{cosv}, is affected by cosmic variance. The inset shows the negligible difference between the standard ``classical'' spectrum (dashed line) and the spectrum with Wheeler--DeWitt quantum corrections (solid line), at $2<\ell<2.5$.}
\end{figure}

As shown in Fig.\ \ref{fig1}, WDW quantum corrections are extremely small even in the most generous estimate, and they are completely drowned by cosmic variance.


\section{Loop quantum cosmology and observations}\label{lqc}


\subsection{The model with inverse-volume corrections}\label{th2}

\subsubsection{Homogeneous background}

Loop quantum gravity is based upon a first-order formulation of gravitational degrees of freedom in terms of the spatial densitized triad field $E^\a_i$ and the Ashtekar--Barbero connection $A_\a^i$, where $i=1,2,3$ is an internal index in the $su(2)$ algebra.
While the connection is not quantized into a well-defined operator, its $SU(2)$-valued holonomy along an edge (with representation defined by the edge spin label $j_e$) is a sensible operator.
Thus, the basic quantities to be quantized are fluxes (integrals of the triad on spatial surfaces) and holonomies. In a (quasi-)FLRW universe, the densitized triad and connection both reduce to one non-trivial component, $E^\a_i=p\delta^\a_i$ and $A_\a^i=c\delta_\a^i$, where $p=a^2$, $c=\gamma\dot{a}$, and elementary edge lengths are all equal to some common value $L=a l_0$. Then, fluxes reduce to $F=l_0^2 p=L^2$ and holonomies along an edge $e$ of comoving length $l_0$ are $h_e=\exp(l_0\tau_ic)= \cos(l_0c/2)+2\tau_i\sin(l_0c/2)$, where $\tau_i=\rmi\s_i/2$ are Pauli matrices. 

We point out that in an exactly FLRW background the universe is perfectly homogeneous and there is no meaningful way to subdivide it into small cells of proper size $L$. Thus, the comoving scale $l_0$ is actually arbitrary and corresponds to the size $\cV_0^{1/3}$ of the fiducial volume in which the Hamiltonian constraint is defined. In this context, inverse-volume corrections depend on an unphysical quantity and should be removed, for instance regarding $\cV_0$ as a regulator and taking the limit $\cV_0\to\infty$. This situation, however, is only a mathematical artifact of the purely homogeneous background, which is not a realistic model of Nature. The full theory does include these corrections.

On the other hand, in the presence of inhomogeneities the lattice picture makes sense (because sub-volumes can be distinguished from one another) and fluxes and holonomies can be defined on each individual cell, not on the overall fiducial volume. The linear scale $L$ is related to the quantum state via its labels $j_e$ and, depending on what spin numbers are realized, it does not need to be exactly the Planck length. Instead of using the $j_e$ and their complicated dynamics (presently not under full control) it is more convenient to adopt $L$ as a phenomenological
parameter. Effective quantum dynamics is then expected to have the cells vary with time. The freedom to choose a global clock in a quasi-homogeneous scenario allows us to pick, e.g., the scale factor $a$ as the time variable, and to regard $L=L(a)$ as time dependent. This is the so-called lattice-refinement picture \cite{lqcr2,bo609,BC}. However unsatisfactory this picture may be ($L$ still contains a high degree of arbitrariness), it allows one to do some phenomenology, with the hope to connect it with the full theory when time is ripe.

A crucial consequence of lattice refinement is that inverse-volume corrections are now phenomenologically meaningful. These quantum corrections arise due to the presence of inverse-volume expressions in the super-Hamiltonian constraint, both in the gravity and matter sector. Inverse volumes (i.e., inverse powers of the determinant of the densitized triad) are not densely defined operators and they must be rexpressed by the so-called ``Thiemann's trick'' in terms of holonomies and positive volume powers, at the classical level before quantizing.

Therefore, contrary to the WDW model, also the background equations of motion (and the slow-roll parameters as well) are deformed by quantum corrections. For a matter scalar field, one has
\bs\ba
&&H^2=\frac{\k^2}{3}\,\a\left[\frac{\dot\phi^2}{2\nu}+V(\phi)\right]\,,\label{frw}\\
&& \ddot\phi+3H\left(1-\frac{\rmd\ln\nu}{\rmd\ln
p}\right)\dot\phi+\nu V_{,\phi}=0\,, \label{kg}
\ea\es
where $\a(a)$ and $\nu(a)$ are inverse-volume corrections in the gravity and matter sector, respectively. Later we shall be interested in the semi-classical limit where quantum corrections are small, in which case
\be\label{qq}
\a(a)= 1+\alpha_0 \dpl(a)\,,\qquad \nu(a)= 1+\nu_0 \dpl(a)\,,
\ee
where $\a_0$ and $\nu_0$ are positive constants (calculable in a pure FLRW case, arbitrary in the lattice refinement picture) and
\be\label{dpl2}
\dpl:= \left[\frac{\lp}{L(a)}\right]^m\propto a^{-\sigma}\,.
\ee
Here $m$ is an $O(1)$ constant dependent on the quantization scheme (e.g., \cite{BC,BCT2}) and $\s$ is determined by a power-law \emph{Ansatz} for the function $L(a)$. While a natural value is $\s=6$ in pure FLRW, in lattice refinement the only constraint is $\s\geq 0$. In general, however, the background inflates only if $\s\lesssim O(1)$ \cite{BC}.

\subsubsection{Perturbations}

To obtain the dynamics of inhomogeneities, we follow the effective constraints method (see \cite{ALS,AKL,WiE1,AAN,FMO,WiE2} for other approaches). 
 The strategy of applying perturbation theory in the classical constraints differs from the one employed in standard cosmology (where the action or the Einstein equations are perturbed), although it was considered in the past \cite{Lan94}.
Perturbing the Ashtekar--Barbero variables, $E_i^\a=p\delta_i^\a+\delta E_i^\a$, $A_\a^i=c\delta_\a^i+\delta A_\a^i$, and imposing commutation relations among the perturbation components, one works out the perturbed form of the seven first-class constraints: the super-Hamiltonian, the three components of the diffeomorphism constraint, and the three components of the Gau\ss\ constraint generating infinitesimal $su(2)$ gauge transformations in the internal space. To capture loop quantum gravity effects, however, one considers effective constraints ${\cal C}_a$ encoding inverse-volume and/or holonomy corrections. For instance, the effective Hamiltonian constraint with inverse-volume corrections is assumed to be
\be
{\cal C}[N]\sim \int \rmd^3 x N [\a(E)\cH_g+\nu(E)\cH_\pi+\varrho(E)\cH_\nabla+\cH_V]\,,
\ee
where $N$ is the lapse function, $\cH_g$, $\cH_\pi$, $\cH_\nabla$ and $\cH_V$ are the contributions of, respectively, gravity, the scalar field momentum, spatial Laplacian and  potential, and $\a$, $\nu$ and $\varrho$ are correction functions (which depend only on the densitized triad \cite{BCT2}). These functions can be taken to be of the form $1+O(\dpl)$ in the semi-classical limit.

Closure of the effective constraint algebra must be imposed for consistency, $\{{\cal C}_a,{\cal C}_b\}=f_{ab}^{\ \ c}(A,E) {\cal C}_c$. The absence of anomalies is guaranteed by introducing counterterms in the algebra (and, hence, in the perturbed equations of motion).
After some early works based on toy models where the constraint algebra was not closed explicitly \cite{ew1,ew2,ew3,ew4,ew5,ew6,ew7}, the full set of constraints with small inverse-volume corrections was derived for vector \cite{BH1}, tensor \cite{BH2}, and scalar modes \cite{BHKS,BHKS2}. The gravitational wave spectrum has been studied in \cite{CMNS2,CH}, while the scalar spectrum and the full set of linear-order cosmological observables were found in \cite{BC}. The observability of and experimental constraints on the quantum corrections were finally considered in \cite{BCT1,BCT2,Cal12}.

In the presence of small inverse-volume corrections, after anomaly cancellation the system of perturbed equations for scalar and tensor modes (vector modes are damped during inflation) reduces exactly to two equations:
\bs\label{mue}\ba
&& u''-\left(s^2_{\rm inv}\Delta+\frac{z_{\rm inv}''}{z_{\rm inv}}\right)u =0\,,\label{mus}\\
&& w''-\left(\a^2\Delta+\frac{\tilde a_{\rm inv}''}{\tilde a_{\rm inv}}\right)w=0\,,
\ea
where primes denote derivatives with respect to conformal time ($'=\p_\tau=a\p_t$), $u=z\cR$ is the Mukhanov--Sasaki variable encoding scalar perturbations, 
\be
z_{\rm inv}:= \frac{\phi'}{H}\left[1+\left(\frac{\a_0}{2}-\nu_0\right)\dpl\right]
\ee
is a background function (quantum corrected as well), $\cR$ is the gauge-invariant comoving curvature perturbation (its LQC expression can be found in \cite{BHKS2,BC}), 
\ba
s^2_{\rm inv} &:=& 1+\chi(\a_0,\nu_0,\s)\dpl\,,\\ 
\chi &:=&\frac{\s\nu_0}{3}\left(\frac{\s}{6}+1\right)+\frac{\a_0}{2}\left(5-\frac{\s}{3}\right)
\ea
is the square propagation speed of the perturbation (discussed in \cite{BC} and positive in all reasonable scenarios), $\Delta$ is the spatial Laplacian, $w=\tilde a_{\rm inv} h$ is the gauge-invariant variable associated with both tensor modes, and
\be
\tilde a_{\rm inv}:= a\left(1-\frac{\a_0}{2}\dpl\right)\,.
\ee\es
The parameter space is extended to include the coefficients appearing in Eqs.\ \Eq{qq} and \Eq{dpl2}. However, self-consistency of the constraint algebra imposes a condition among $\a_0$, $\nu_0$ and $\s$, thus making one of them dependent \cite{BC}:
\be\label{extra}
\a_0\left(\frac{\s}6-1\right)-\nu_0\left(\frac{\s}6+1\right)\left(\frac{\s}3-1\right)=0\,.
\ee
The fact that scalar perturbations reduce to just one degree of freedom $u$ obeying a closed equation is related to conservation of $\cR$ at large scales \cite{BC}. Failure of closing the algebra exactly would immediately spoil also this property.

The somewhat unexpected possibility that LQC quantum corrections be large even during inflation is a reflection of the way these corrections enter the physics: The structure of spacetime itself is deformed by quantum effects, via the effective constraints. The theory is diffeomorphism invariant, but not with respect to the standard classical transformations. Gauge transformations belonging to a deformed algebra no longer correspond to ordinary coordinate transformations on a manifold. Thus, in order to take the new gauge structure into account one should rely only on gauge-invariant perturbations. This philosophy (first quantize the classical system, then cast it in gauge-invariant variables) is embodied in the Mukhanov equations \Eq{mue}. 

One might wonder whether one would get the same results by fixing the gauge \emph{before} quantizing. However, gauge fixing and quantization do not commute because the latter deeply affects the very notion of gauge invariance. Whenever gauge-ready variables can be constructed after quantizing, the gauge-invariant approach must be preferred. The price to pay in doing otherwise is, in the least conservative interpretation, to produce unphysical perturbative modes (this may happen also in standard cosmology, due to an illegal choice of gauge \cite{MFB}) or, more conservatively, to obtain an incomplete version of the perturbed quantum equations which, at best, can be interpreted as a physically different quantum system. Also ignoring backreaction of the metric and considering just a perturbed test scalar is undesirable, contrary to the WDW case, because backreaction contributes to the actual form of quantum gauge transformations and hence of the gauge-invariant variables. Again, this can lead to an incomplete treatment in partial disagreement with the full gauge-invariant equations.

\subsubsection{Observables}

The scalar spectrum is the expectation value of $\cR$ over a momentum ensemble at large scales, evaluated at horizon crossing:
\be
{\cal P}_{\rm s} \equiv \frac{k^3}{2\pi^2z^2_{\rm inv}} \left\langle |u_{k\ll k_*}|^2\right\rangle\Big|_{k=k_*}\,.
\ee
Solving the Mukhanov equation \Eq{mus} asymptotically and plugging the solution in the above formula, one obtains the LQC version of Eq.\ \Eq{cP10wdw} with
\be 
C_k^2\approx 1+\g_{\rm s}\dpl\,,\qquad \g_{\rm s}:= \nu_0\left(\frac{\s}{6}+1\right)+\frac{\s\a_0}{2\e}-\frac{\chi}{\s+1}\,.
\ee
In the limit case $\s\to 0$, the quantum correction is constant and the only change with respect to the classical case is the normalization of the spectrum. Then, $\g_{\rm s}=\nu_0-5\a_0/2$ could be of either sign. If $\s\neq 0$, there is a large-scale \emph{enhancement} of power because $\dpl\sim k^{-\s}$ at horizon crossing and $\g_{\rm s}>0$ due to the dominating term $\propto \e^{-1}$. Similarly, the scalar index is
\be
n_{\rm s}-1 \approx 2\eta-4\e+\s\g_{n_{\rm s}}\dpl\,,\qquad \g_{n_{\rm s}}:= \a_0-2\nu_0+\frac{\chi}{\s+1}\,,\label{nslqc}
\ee
while the scalar running reads
\be
\a_{\rm s} \approx 2(5\e\eta-4\e^2-\xi^2)+\s(4\tilde\e-\s \g_{n_{\rm s}})\dpl\,,
\ee
where $\tilde\e:= \a_0({\s}/{2}+2\e-\eta)+\nu_0({\s}/6-1)\e$.

Due to the possibly large size of the quantum corrections, it will be useful to complete the set of first-order observables and include also the tensor sector. The gravitational spectrum is
\be
\cP_{\rm t} := \frac{32 G}{\pi}\frac{k^3}{\tilde a_{\rm inv}^2} \left\langle |w_{k\ll k_*}|^2\right\rangle\big|_{k=k_*}\,,
\ee
leading to \cite{BC,CH}
\be
\cP_{\rm t} \approx 64\pi G\left(\frac{H}{2\pi}\right)^2
\left(1+\g_{\rm t}\dpl\right)\,,\qquad \g_{\rm t}:= \frac{\s-1}{\s+1}\a_0\,.\label{tesp}
\ee
The tensor index is
\be
n_{\rm t}:=\frac{\rmd\ln {\cal P}_{\rm t}}{\rmd\ln k}\approx -2\e-\s\g_{\rm t}\dpl\,.\label{nt}
\ee
Finally, the tensor-to-scalar ratio is
\be
r:= \frac{{\cal P}_{\rm t}}{{\cal P}_{\rm s}}\approx 16\e[1+(\g_{\rm t}-\g_{\rm s})\dpl]\,,
\ee
which yields the consistency relation
\be\label{tts2}
r = -8\{n_{\rm t}+[n_{\rm t}(\g_{\rm t}-\g_{\rm s})+\s\g_{\rm t}]\dpl\}\,,
\ee
to be plugged into numerical codes in the place of the classical one $r=-8n_{\rm t}$.


\subsection{Experimental bounds}\label{ex2}

Because of the delicate interplay between quantum corrections and the requirement of intersecting the allowed windows in the parameter space in a common consistent region, the possibility clearly arises that this model of loop quantum cosmology be falsifiable by near-future observations. The present status at least provides stringent bounds on quantum corrections.

As in section \ref{ex1}, one rewrites the observables in terms of the potential-dependent slow-roll parameters \Eq{vsr}; the resulting lengthy expressions can be found in \cite{BCT2}. Since
\be\label{aln}
\alpha_{\rm s}^{(m)} (k_0) \approx (-1)^m \s^{m-1}f_{\rm s}\dpl (k_0)\,,
\ee
where
\be
f_{\rm s}:= \frac{\sigma [3\a_0(13\s-3)+ \nu_0\s(6+11\s)]}{18(\sigma+1)}\,,
\ee
the scalar spectrum \Eq{Ps2} becomes
\ba
{\cal P}_{\rm s} (k)&\approx&{\cal P}_{\rm s} (k_0) \exp \left\{ [n_{\rm s}(k_0)-1]x
+\frac{\alpha_{\rm s}(k_0)}{2}
x^2\right.\nonumber\\
&&\left.+f_{\rm s} \dpl(k_0) \left[x \left(1-\frac12 \sigma x \right)
+\frac{1}{\sigma} (e^{-\sigma x}-1) \right] \right\}\,.
\label{Psfinal}
\ea
This is the expression to be used in numerical analyses and when comparing the LQC signal with cosmic variance. 

Before doing so, we notice the existence of a theoretical upper bound on the quantum correction $\de_{\textsc{lqc}}:=\a_0\dpl$. (Equation \Eq{extra} allows to remove $\nu_0$ from parameter space, except in the case $\s=3$ which can be treated separately.) For the validity of the linear expansion of the perturbation formul\ae\ where the $O(\dpl)$ truncation has been 
systematically implemented, we require that
\be
\de_{\textsc{lqc}}(k)=\de_{\textsc{lqc}}(k_0) \left(\frac{k_0}{k}\right)^{\sigma}=\de_{\textsc{lqc}}(k_0) \left(\frac{\ell_0}{\ell}\right)^{\sigma}<1
\label{deltak}
\ee
for all wavenumbers relevant to the CMB anisotropies. For the pivot scale $\ell_0=29$, the quadrupole $\ell=2$ gives the bound $\de_{\textsc{lqc}}(k_0)<\de_{\textsc{lqc}}^{\rm max}=14.5^{-\sigma}$, shown in Table \ref{tab1} for some choices of $\s$.
\begin{table*}
\begin{center}
\begin{andptabular}{c|cccccc}{\label{tab1} \twocolcaption
Theoretical priors on the upper bound $\de_{\textsc{lqc}}^{\rm max}$
and 95\% C.L.\ upper limits of $\de_{\textsc{lqc}}=\a_0\dpl$ constrained by observations for a quadratic potential with different values of $\sigma$ and at the pivot scale $k_0=0.002\,{\rm Mpc}^{-1}$  \cite{BCT2}. The likelihood analysis is omitted for $\s=6$  since the signal is below the cosmic variance threshold already when $\s=2$. For $\s=3$, the parameter $\delta_{\textsc{lqc}}=\nu_0\dpl$ has been used.}%
$\sigma$ & 0.5 & 1 & 1.5 & 2 & 3 & 6  \\\hline
$\de_{\textsc{lqc}}^{\rm max}$ & 0.26 & $6.9 \times 10^{-2}$ & $1.8 \times 10^{-2}$ 
& $4.7 \times 10^{-3}$ & $3.2 \times 10^{-4}$ & $1.0 \times 10^{-7}$ \\
$\de_{\textsc{lqc}}$ & 0.27 & $3.5 \times 10^{-2}$ & $1.7 \times 10^{-3}$ & $6.8 \times 10^{-5}$ 
& $4.3 \times 10^{-7}$ & -- \\\hline
\end{andptabular}
\end{center}
\end{table*}

\begin{figure}
\centering
\includegraphics[width=7cm]{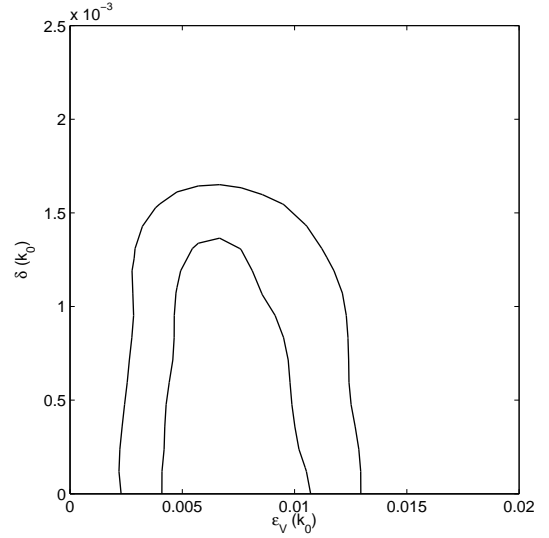}
\caption{Two-dimensional marginalized distribution for the inverse-volume LQC quantum 
correction $\de_{\textsc{lqc}}(k_0)$ and the slow-roll parameter $\e_\V (k_0)$ 
with the pivot $k_0=0.002$ Mpc$^{-1}$ for $\s=1.5$ and a quadratic potential,
constrained by the joint data analysis of WMAP7, LSS (including BAO), HST, SN Ia, and BBN. 
The internal and external lines correspond to the 68\% and 95\% 
confidence level, respectively \cite{BCT2}.\label{fig2}}
\end{figure}

To illustrate some of the possibilities CMB data manipulations can offer to constrain quantum gravity models with free parameters, we recall the likelihood analysis carried out in \cite{BCT2,BCT1} for the quadratic potential (among others). The Cosmological Monte Carlo (\textsc{CosmoMC}) code \cite{Antony} was run with the data of WMAP7 \cite{Kom10} combined with large-scale structure (LSS) \cite{Reid} (including BAO), HST \cite{HST}, Supernovae type Ia (SN Ia) \cite{SNIa}, and Big Bang Nucleosynthesis (BBN) \cite{BBN}, assuming a $\Lambda$CDM model. Figure \ref{fig2} shows an example of likelihood profile for $\s=3/2$ in the plane $(\e_\V,\de_{\textsc{lqc}})$. Both parameters are evaluated at the pivot scale $k_0=0.002$ Mpc$^{-1}$. Obviously, negligible or exactly vanishing quantum corrections are compatible with observations.  On the other hand, from the 95\% confidence-level contour one sees that quantum corrections above  $\de_{\textsc{lqc}}(k_0)\gtrsim 1.7 \times 10^{-3}$ can be excluded. This and the upper bounds for various $\s$'s are reported in Table \ref{tab1}. Except for extreme values $\s\ll 1$, the observational upper bounds are consistent with the theoretical prior, thus verifying an important internal check of the model.

\begin{figure}
\centering
\includegraphics[width=\columnwidth]{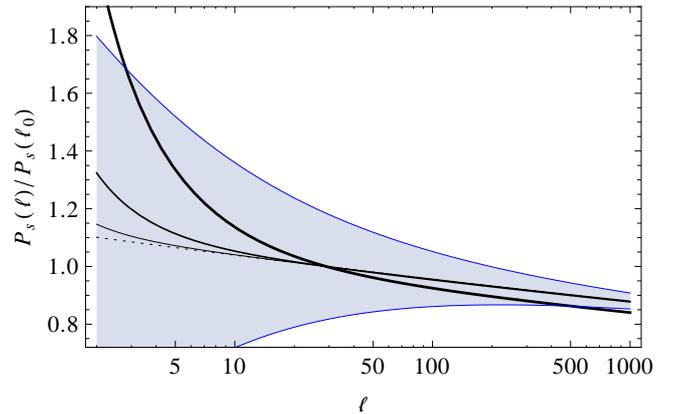}
\caption{Log-linear plot of the LQC primordial scalar spectrum ${\cal P}_{\rm s}(\ell)$ with inverse-volume quantum corrections for a quadratic inflaton potential, with $\epsilon_{\V}(k_0)=0.009$ and for the pivot wavenumber $k_0=0.002$ Mpc$^{-1}$, corresponding to $\ell_0= 29$.
The classical case is represented by the dotted line, while solid curves correspond to  $\s=1,1.5,2$ (decreasing thickness). The shaded region is affected by cosmic variance. \label{fig3}}
\end{figure}

Comparing the table entries with the upper bound for the WDW quantum correction, Eq.\ \Eq{dwdw}, we see that LQC inverse-volume corrections can be orders of magnitude larger when $\s\lesssim 2$. The scalar power spectrum for various values of $\s$ is shown in Fig.\ \ref{fig3} against  cosmic variance. When $\s\lesssim 1$, quantum corrections are strong enough to overcome the error from cosmic variance. Whether these parameter values are realistic in a more complete theory remains, however, to be seen.


\subsection{The model with holonomy corrections}\label{th3}

Another type of quantum effect in the dynamics, holonomy corrections, is realized in a highly non-linear fashion (by 
construction, from the exponentiation $h_e$ of curvature components) and it becomes important when the Hubble radius is 
about the size of the lattice scale, $H^{-1}\sim L$. From the classical Friedmann equation $H^2=8\pi G\rho/3$, this 
regime heuristically defines the critical energy density \Eq{rhoqg} and the holonomy correction
\be
\delta_{\rm hol}:= \frac{\rho}{\rho_\textsc{qg}}\,. 
\ee
The homogeneous background is modified accordingly. While Eq.\ \Eq{kg} remains the same, the Friedmann equation \Eq{frw} is further corrected as
\be\label{frwlqc}
H^2=\frac{\k^2}{3}\,\rho(\a-\delta_{\rm hol})\,.
\ee
Crucially, the Hubble parameter is not simply $H=\dot a/a$ but the ``polymeric'' expression
\be\label{hub}
H = \frac{\sin [2\g L (\dot a/a)]}{2\g L}\,.
\ee
Even in a perfectly homogeneous background, $\rho_\textsc{qg}$ is not constant except for a specific choice of quantum ambiguity parameters, such that the elementary closed-holonomy area coincides with the Planck area $L^2\propto \lp^2$ (``improved dynamics''  \cite{CQC,lqcr1,lqcr2}). For this choice, and ignoring or removing inverse-volume corrections ($\a=1$), the right-hand side of Eq.\ \Eq{frwlqc} vanishes at $\rho=\rho_\textsc{qg}$, where the Hubble parameter $H\to 0$ and the big-bang singularity of classical cosmology is replaced by a bounce.

There are indications that holonomy corrections are not significant in the energy regime of inflation, but only at 
near-Planckian densities \cite{APSII}. This is suggested by effective equations for certain matter contents with a 
dominating kinetic energy \cite{Boj74,BNMR}. Another argument is the following \cite{BCT2}. Inverse-volume and holonomy corrections are related to each other by
\be
\dpl= \left(\frac{8\pi G}{3}\rho_\textsc{qg} \lp^2\right)^{\frac{m}{2}}\propto
 \left(\frac{\rho_\textsc{qg}}{\rho_{\rm Pl}}\right)^{\frac{m}{2}}= \left(\frac{\rho}{\rho_{\rm Pl}} \delta_{\rm 
hol}^{-1}\right)^{\frac{m}{2}}\,. \label{dpl}
\ee
Inverse-volume corrections are sizable when the quan-tum-gravity density (not the inflationary one) is close to the 
Planck density. They can be still large at small energy densities, where however holonomy corrections are small. Thus, as the energy density decreases in an expanding universe there is a competition of the relative size of inverse-volume and holonomy corrections, the latter falling to small values when the former can be still large. For instance, in the inflationary regime \Eq{wdwqc} and for the typical value $m=4$ Eq.\ \Eq{dpl} yields $\delta_{\rm hol} \sim 10^{-8}/\sqrt{\dpl}$, and having small holonomy corrections of size $\delta_{\rm hol}<10^{-6}$ would require inverse-volume corrections larger than $\dpl>10^{-4}$.

This argument is only heuristic and a full cosmological analysis is required to settle the issue. This is now at hand because perturbation theory has been worked out already. In fact, the closure of the constraint algebra has been verified also in the presence of holonomy corrections for vector and tensor modes \cite{BH1,BH2,MCBG}, as well as in the scalar sector \cite{CMBG,CaB,CBGV}. Just as in the case of inverse-volume corrections, the constraint algebra is deformed by quantum effects and gauge transformations do not correspond to standard diffeomorphisms.
Notice that the lattice refinement interpretation also affects holonomy corrections, since they feature the same 
phenomenological parameter $L$ as inverse-volume corrections. The Mukhanov equations for scalar and tensor modes are \cite{BH2,CMBG,CBGV}
\bs\label{mue2}\ba
&& u''-\left(s^2_{\rm hol}\Delta+\frac{z_{\rm hol}''}{z_{\rm hol}}\right)u =0\,,\label{mus2}\\
&& w''-\left(s^2_{\rm hol}\Delta+\frac{\tilde a_{\rm hol}''}{\tilde a_{\rm hol}}\right)w=0\,,
\ea
where the effective propagation speed and background funcion $z_{\rm hol}$ and $\tilde a_{\rm hol}$ read
\ba
&& s^2_{\rm hol} := \cos[2\g L (\dot a/a)]=1-2\delta_{\rm hol}\,,\\
&& z_{\rm hol} := \frac{\phi'}{H}\,,\qquad \tilde a_{\rm hol}:=\frac{a}{|s_{\rm hol}|} \,,
\ea\es
and $H$ is given by Eq.\ \Eq{hub}. These expressions should be compared with their inverse-volume counterparts \Eq{mue}. The propagation speed is never super-luminal ($|s^2_{\rm hol}|\leq 1$), but it does change sign near the bounce. This marks a possible instability, or even a change of effective spacetime signature at near-Planckian scales\cite{BoP}, in a super-inflationary early era. The physical significance of these features is still under inspection.

Cosmological observational signatures of holonomy effects have been studied for the tensor sector alone 
\cite{CMNS2,Mie1,GB1,GB2,GB3}. For this reason, we do not yet have a detailed comparison with experiments as in the WDW and inverse-volume LQC cases. With respect to the inverse-volume case, the analysis of the spectra is complicated by the analytic form of holonomy corrections. In general, tensor modes are amplified during the bounce. However, after the bounce these modes are enhanced by inflationary expansion later than in the classical case, and the spectrum is thus \emph{suppressed} at low multipoles, as \cite{GB2,GB3}
\be
\cP_{\rm t}\propto k^2\,\qquad k\to 0\,,
\ee
on a de Sitter background. It also shows an oscillatory pattern, progressively damped towards small scales. 
The gravitational spectrum is notoriously difficult to detect by itself, and information from the scalar spectrum (which, from Eq.\ \Eq{mus2}, is expected to behave similarly to the tensor one) will be needed, also to determine whether the large-scale suppression is beyond the cosmic-variance noise and therefore observable.


\section{Non-Gaussianity}\label{nonga}

The effect of quantum corrections goes beyond linear perturbation theory and higher-order observables can be calculated. As the perturbative level increases, the statistics of inhomogeneous fluctuations deviates from the Gaussian one and odd-order correlation functions acquire non-vanishing values. In particular, the bispectrum (three-point correlation function of the curvature perturbation) can be constrained by observations. 

To the best of our knowledge, there is only one work on inflationary non-Gaussianity in loop quantum cosmology with inverse-volume corrections \cite{LCGH}, and none in the WDW case. A detailed calculation of the momentum-dependent bispectrum shows that no appreciable LQC signal can be detected. We can in fact reach the same conclusion here by a model-independent shortcut, valid only in the so-called squeezed limit (constant non-linear parameter) but beyond perturbation theory and both for LQC and WDW quantum cosmology.

Let $\zeta$ be the curvature perturbation on uniform density hypersurfaces. The latter is a gauge-invariant quantity proportional to the comoving curvature perturbation $\cR$ in standard inflation; their relation in the presence of inverse-volume corrections has not been studied yet, but what follows is fairly independent on this detail. In momentum space, the three-point correlation function of $\zeta$ is
\be\label{defB}
\left\langle\zeta_{{\bf k}_1}\zeta_{{\bf k}_2}\zeta_{{\bf k}_3}\right\rangle=:(2\pi)^3\delta({\bf k}_1+{\bf k}_2+{\bf k}_3)B_\zeta(k_1,k_2,k_3)\,,
\ee
where $B_\zeta$, called bispectrum, is defined by 
\be
B_\zeta(k_1,k_2,k_3)=\frac65 f_{\rm NL}(k_1,k_2,k_3) \sum_{\a<\b}P_\zeta(k_\a)P_\zeta(k_\b)\,,\label{bisp}
\ee
where $\a,\b=1,2,3$, $f_{\rm NL}$ is called non-linear parameter and is momentum dependent in general, and $P_\zeta$ is the spectrum of $\zeta$. The form of the non-linear parameter depends on the model of primordial perturbations. In the simplest case \cite{GLMM,VWHK,KoS}, one decomposes the non-linear curvature perturbation $\zeta_{\rm NL}({\bf x})$ into a Gaussian linear part $\zeta$ and a non-linear part:
\be\label{local}
\zeta_{\rm NL}=\zeta+\zeta^{\rm N}=\zeta+\frac{3}{5}f^{\rm local}_{\rm NL}\left(\zeta^2-\langle\zeta^2\rangle\right)\,,
\ee
where the non-linear parameter $f_{\rm NL}^{\rm local}$ is constant. By definition, $\langle\zeta_{\rm NL}\rangle=\langle\zeta\rangle=0$. Then, a direct calculation of the bispectrum shows that
\be\label{fnlkn}
f_{\rm NL}(k_1,k_2,k_3)=f_{\rm NL}^{\rm local}\,.
\ee
In fact, the Fourier transform of the non-linear part is
\be
\zeta_{\bf k}^{\rm N}=\frac{3}{5}f_{\rm NL}^{\rm local}\left[-(2\pi)^3\delta({\bf k})\langle\zeta^2\rangle+\int\frac{\rmd^3{\bf p}}{(2\pi)^3}\zeta_{\bf p}\zeta_{{\bf p}-{\bf k}}\right]\,.
\ee
The first term stems from the fact that the auto-correlat-ion function is ${\bf x}$ independent. Since all momenta must not vanish at the same time, this piece can be thrown away. The second term enters into the three-point function, which at lowest order is (e.g., \cite{CQC})
\ba
\langle\zeta_{{\bf k}_1}^{\rm NL}\zeta_{{\bf k}_2}^{\rm NL}\zeta_{{\bf k}_3}^{\rm NL}\rangle&\approx&
\langle\zeta_{{\bf k}_1}\zeta_{{\bf k}_2}\zeta_{{\bf k}_3}^{\rm N}\rangle+({\bf k}_3\leftrightarrow{\bf k}_2)+({\bf k}_3\leftrightarrow{\bf k}_1)\nonumber\\
&=&(2\pi)^3\delta({\bf k}_1+{\bf k}_2-{\bf k}_3) \frac{3}{5}f_{\rm NL}^{\rm local}2P_\zeta(k_1)P_\zeta(k_2)\nonumber\\
&&+({\bf k}_3\leftrightarrow{\bf k}_2)+({\bf k}_3\leftrightarrow{\bf k}_1)\,,\label{zzz}
\ea
which yields Eq.~\Eq{fnlkn} after comparing Eqs.~\Eq{defB} and \Eq{bisp}. The decomposition \Eq{local} is pointwise in configuration space and for this reason it is called \emph{local model}. For a power-law scalar spectrum $\cP_{\rm s}\propto k^{n_{\rm s}-1}$, the local bispectrum reads
\be\label{locB}
B_\zeta^{\rm local}(k_1,k_2,k_3)=\frac{6}{5}f_{\rm NL}^{\rm local}A_\zeta^2\sum_{\a<\b}\frac{1}{(k_\a k_\b)^{4-n_{\rm s}}}\,,
\ee
where $A_\zeta$ is a constant amplitude. This expression can be converted into one with spherical multipoles.

The expression \Eq{locB} peaks at the \emph{squeezed} limit where one of the edges of the triangle $({\bf k}_1,{\bf k}_2,{\bf k}_3)$ collapses \cite{Mal02,BCZ}:
\be\label{squee}
k_1\approx k_2\gg k_3\,,\qquad k_3\approx 0\,.
\ee
Sending, e.g., $k_3\to 0$, by conservation of momenta one has ${\bf k}_1\sim -{\bf k}_2$ and
\be\label{Bsq}
B_\zeta^{\rm local}(k_1,k_1,k_3\to 0)=\frac{12}{5}f_{\rm NL}^{\rm local}P_\zeta(k_1)P_\zeta(k_3)\,.
\ee
Measuring the bispectrum in this configuration, one can obtain an estimate of $f_{\rm NL}^{\rm local}$. In the local bispectrum, small- and large-scale modes are coupled together.

The squeezed limit can be understood in a fairly intuitive way in all models where the curvature perturbation $\zeta$ is constant at large scales \cite{Mal02,CrZ}. Split $\zeta$ into a corse-grained and a fine-grained perturbation,
\ba
\zeta(\tau,{\bf x}) &=&\int_{k<k_*}\frac{\rmd^3{\bf k}}{(2\pi)^3}\zeta_{\bf k}(\tau)\rme^{\rmi {\bf k}\cdot {\bf x}}+\int_{k>k_*}\frac{\rmd^3{\bf k}}{(2\pi)^3}\zeta_{\bf k}(\tau)\rme^{\rmi {\bf k}\cdot {\bf x}}\nonumber\\
&=:& \zeta_{\rm c}(\tau,{\bf x})+\zeta_{\rm q}(\tau,{\bf x})\,.\label{cq}
\ea
In the limit \Eq{squee}, $\zeta_{{\bf k}_3}$ is larger than the Hubble horizon and can be treated as constant in time. Then $\zeta({\bf x}_3)\sim\zeta_{\rm c}({\bf x}_3)$ defines a new coordinate background ${\bf x}'\approx[1+\zeta_{\rm c}({\bf x}_3)]{\bf x}$ inside the horizon. In the new coordinates and up to linear order,
\ba
\zeta_{\rm q}({\bf x}')&\approx&\zeta_{\rm q}({\bf x})+({\bf x}'-{\bf x})\cdot\frac{\rmd}{\rmd{\bf x}}\zeta_{\rm q}({\bf x})\nonumber\\
&\approx&\zeta_{\rm q}({\bf x})+\zeta_{\rm c}({\bf x}_3)\,{\bf x}\cdot\frac{\rmd}{\rmd{\bf x}}\zeta_{\rm q}({\bf x})\,.\label{grz}
\ea
If the linear perturbation $\zeta_{\rm q}({\bf x})$ is Gaussian, in the squeezed limit we have
\ba
\left\langle\zeta({\bf x}_1)\zeta({\bf x}_2)\zeta({\bf x}_3)\right\rangle &\sim& \left\langle\zeta_{\rm q}({\bf x}_1')\zeta_{\rm q}({\bf x}_1')\zeta_{\rm c}({\bf x}_3)\right\rangle\nonumber\\
&\approx&\left\langle\zeta_{\rm c}^2({\bf x}_3)
{\bf x}_1\cdot\frac{\rmd}{\rmd{\bf x}_1}\left[\zeta_{\rm q}({\bf x}_1)\zeta_{\rm q}({\bf x}_2)\right]\right\rangle\nonumber\\
&\approx&\left\langle\zeta_{\rm c}^2({\bf x}_3)\right\rangle_{\rm c}{\bf x}_1\cdot\frac{\rmd}{\rmd{\bf x}_1}\left\langle\zeta_{\rm q}({\bf x}_1)\zeta_{\rm q}({\bf x}_2)\right\rangle_{\rm q}\nonumber\\
&=& \xi_2^{(\zeta)}(0)\frac{\rmd}{\rmd\ln\vr}\xi_2^{(\zeta)}(\vr)\,,
\ea
where in the second line we exploited translation invariance, in the last line we used $\vr=|{\bf x}_1-{\bf x}_2|$ and $\p \vr/\p{\bf x}_1={\bf x}_1/\vr$, and we denoted with $\xi_2^{(\zeta)}$ the two-point correlation functions of $\zeta$. Since the latter goes as $\xi_2^{(\zeta)}\propto \vr^{-(n_{\rm s}-1)}$, one gets
\be
\left\langle\zeta({\bf x}_1)\zeta({\bf x}_2)\zeta({\bf x}_3)\right\rangle\approx -(n_{\rm s}-1)\xi_2^{(\zeta)}(0)\xi_2^{(\zeta)}(\vr)\,.
\ee
Comparing this expression with Eq.~\Eq{Bsq}, we finally obtain
\be\label{sqfnl}
f_{\rm NL}^{\rm local}\approx \frac{5}{12}(1-n_{\rm s})\,.
\ee

For spectra which are almost scale-invariant ($n_{\rm s}-1$ small) at large scales, the level of non-Gaussianity is very low, $f_{\rm NL}\ll1$. Tensor modes produce an even lower signal. This result \cite{Mal02,CrZ} is general enough to be applied both to WDW and loop quantum cosmology, which we have seen to be compatible with almost scale invariance. Therefore, considering the current 95\% C.L.\ bound on the local non-linear parameter coming from combined CMB and large-scale structure \cite{SHSHP} observations, $-5<f_{\rm NL}^{\rm local}<59$ \cite{Kom10}, the non-linear parameter in the squeezed limit is small and the quantum corrections considered here have no appreciable impact on the bispectrum.


\section{Outlook}

Quantum gravitational effects modify the spectra of cosmological perturbations and their imprint in the cosmic microwave background. In this paper, we compared two canonical approaches, the one based on the usual Wheeler--DeWitt quantization and loop quantum cosmology. Wheeler--DeWitt quantum corrections are too small to be detected, even in the most optimistic upper bound, Eq.\ \Eq{dwdw}. The model therefore is not falsifiable, at least under the assumptions made in the derivation of the results, but at least it is compatible with what we observe.

In contrast, LQC inverse-volume corrections can be of much greater size and produce an enhancement, rather than suppression, of the large-scale spectra. While in the WDW case quantum corrections change the inhomogeneous dynamics but leave homogeneous background equations unmodified, in LQC the latter are deformed, too. However, this is not the reason why LQC effects are potentially several orders of magnitude larger than the WDW quantization. Rather, the key ingredient is the scale compared with the Planck energy density $\rho_{\rm Pl}$ in the ratio defining the quantum correction: for WDW it is the inflationary scale $\rho_{\rm infl}$, for LQC it is determined by the characteristic discreteness scale of the semi-classical state describing the quantum universe. This effective energy density can be as large as the Planck density, $\rho_{\rm infl}\ll\rho_\textsc{qg}\lesssim\rho_{\rm Pl}$. 

This also highlights the different origin of the observational bounds presented above. While the WDW quantum correction \Eq{dwdw} is constrained somewhat indirectly via the usual bounds on the inflationary energy scale, in LQC we have some free parameters on which we have little control theoretically, due to the formidable (and yet unsurmounted) difficulties in explicit constructions of cosmological semi-classical states in the full theory. LQC inverse-volume corrections depend on a phenomenological quantum-gravity scale as well as on partly heuristic, partly quantitative arguments indicating how to implement discrete quantum geometry in a quasi-homogeneous cosmological setting. A multi-variate likelihood analysis involving all the cosmological parameters, including LQC ones, is thus more adequate to the task.

Observations constrain LQC inverse-volume quantum corrections below their theoretical upper bound, but in some instances the signal is above the threshold of cosmic variance. Experiments such as PLANCK or of the next generations should then be able to reach the sensitivity to detect a quantum gravity signal or, in its absence, place yet more stringent constraints. In turn, pressure from actual data will stimulate the quest for a better understanding of the fundamental properties of the states of the full theory, and a greater control over parameters which, as the discreteness scale $L$, are presently phenomenological.



\subsection*{Acknowledgments}

The author thanks Martin Bojowald and Claus Kiefer for useful discussions.

\bibliographystyle{andp2012}

\begin{thebibliography}{00}
{\footnotesize

\othercit  
\bibitem{Ori09} \textsc{D.~Oriti} (ed.), Approaches to Quantum Gravity (Cambridge University Press, Cambridge, 2009).

\othercit 
\bibitem{Kie07} \textsc{C.~Kiefer}, Quantum Gravity (Oxford University Press, Oxford, 2012).

\bibitem{Boj11} \textsc{M.~Bojowald}, 
  \doi{10.1007/978-1-4419-8276-6}{\jr{Lect.\ Notes Phys.}  {\bf 835}, 1 (2011)}.

\othercit  
\bibitem{Wil95} \textsc{D.\,L.~Wiltshire}, 
in: Cosmology: The Physics of the Universe, edited by B.~Robson, N.~Visvanathan, and W.\,S.~Woolcock (World Scientific, Singapore, 1996) [\arX{gr-qc/0101003}].

\othercit
\bibitem{CQC} \textsc{G.~Calcagni}, Classical and Quantum Cosmology, in preparation.  

\bibitem{Boj08} M.~Bojowald, 
  \href{http://www.livingreviews.org/lrr-2008-4}{\cob \jr{Living Rev.\ Relativity} {\bf 11}, 4 (2008)}. 
  
\bibitem{lqcr1} \textsc{A.\ Ashtekar} and \textsc{P.\ Singh}, 
 \doi{10.1088/0264-9381/28/21/213001}{\jr{Class.\ Quantum Grav.} {\bf 28}, 213001 (2011)} [\arX{1108.0893}].
 
\bibitem{lqcr2} \textsc{K.\ Banerjee}, \textsc{G.\ Calcagni}, and \textsc{M.\ Mart\'{\i}n-Benito}, 
 \doi{10.3842/SIGMA.2012.016}{\jr{SIGMA} {\bf 8}, 016 (2012)} [\arX{1109.6801}].
 
\bibitem{Bo06b} \textsc{M.\ Bojowald}, \doi{10.1142/S0218271806008942}{\jr{Int.\ J.\ Mod.\ Phys.\ D} {\bf 15}, 1545 (2006)} [\arX{gr-qc/0607130}].

\bibitem{Ash08} \textsc{A.\ Ashtekar},  
  \doi{10.1007/978-90-481-3475-5_7}{\jr{Fund.\ Theories Phys.} {\bf 165}, 163 (2010)} [\arX{0810.0514}].

%
\bibitem{ReS} M.~Reuter and F.~Saueressig, 
  \doi{10.1007/978-3-642-33036-0_8}{\jr{Lect.\ Notes Phys.} {\bf 863}, 185 (2013)} [\arX{1205.5431}].
\bibitem{Gor13} A.~G\"orlich, 
  \doi{10.1007/978-3-642-33036-0_5}{\jr{Lect.\ Notes Phys.} {\bf 863}, 93 (2013)}.

\bibitem{BM}  C.P.\ Burgess and L.\ McAllister, 
\doi{10.1088/0264-9381/28/20/204002}{\jr{Class.\ Quantum Grav.} {\bf 28}, 204002 (2011)} [\arX{1108.2660}]. 
\bibitem{MuW} D.J.~Mulryne and J.~Ward, 
  \doi{10.1088/0264-9381/28/20/204010}{\jr{Class.\ Quantum Grav.} {\bf 28}, 204010 (2011)} [\arX{1105.5421}]. 
\bibitem{Kal07} R.~Kallosh, 
  \doi{10.1007/978-3-540-74353-8_4}{\jr{Lect.\ Notes Phys.} {\bf 738}, 119 (2008)} [\arX{hep-th/0702059}].
\bibitem{CQ}  M.~Cicoli and F.~Quevedo, 
 \doi{10.1088/0264-9381/28/20/204001}{\jr{Class.\ Quantum Grav.} {\bf 28}, 204001 (2011)} [\arX{1108.2659}]. 
\bibitem{CPV} E.J.~Copeland, L.~Pogosian, and T.~Vachaspati, 
  \doi{10.1088/0264-9381/28/20/204009}{\jr{Class.\ Quantum Grav.} {\bf 28}, 204009 (2011)} [\arX{1105.0207}]. 
\bibitem{Avg11} A.~Avgoustidis, E.J.~Copeland, A.~Moss, L.~Pogosian, A.~Pourtsidou, and D.A.~Steer,
  \doi{10.1103/PhysRevLett.107.121301}{\jr{Phys.\ Rev.\ Lett.} {\bf 107}, 121301 (2011)} [\arX{1105.6198}].
\bibitem{Bra11} R.H.~Brandenberger, 
  \doi{10.1088/0264-9381/28/20/204005}{\jr{Class.\ Quantum Grav.} {\bf 28}, 204005 (2011)} [\arX{1105.3247}].
\bibitem{Bra12} R.H.~Brandenberger,   
  \doi{10.1007/978-3-642-33036-0_12}{\jr{Lect. Notes Phys.} {\bf 863}, 333 (2013)} [\arX{1203.6698}].
\bibitem{MaK} R.\ Maartens and K.\ Koyama, \href{http://www.livingreviews.org/lrr-2010-5}{\cob \jr{Living Rev.\ Relativity} {\bf 13}, 5 (2010)}. 
\bibitem{Leh08} J.-L.~Lehners, 
  \doi{10.1016/j.physrep.2008.06.001}{\jr{Phys.\ Rept.} {\bf 465}, 223 (2008)}
  
   [\arX{0806.1245}].
\bibitem{Leh11} J.-L.~Lehners, 
  \doi{10.1088/0264-9381/28/20/204004}{\jr{Class.\ Quantum Grav.} {\bf 28}, 204004 (2011)} [\arX{1106.0172}].

\othercit
\bibitem{HTe}   \textsc{M.~Henneaux} and \textsc{C.~Teitelboim}, Quantization of Gauge Systems (Princeton University Press, Princeton, 1994).

\bibitem{CGO2}  \textsc{G.~Calcagni}, \textsc{S.~Gielen}, and \textsc{D.~Oriti},
  \doi{10.1088/0264-9381/29/10/105005}{\jr{Class.\ Quantum Grav.} {\bf 29}, 105005 (2012)}
 [\arX{1201.4151}].

\bibitem{HaH}   \textsc{J.\,J.~Halliwell} and \textsc{S.\,W.~Hawking},
  \doi{10.1103/PhysRevD.31.1777}{\jr{Phys.\ Rev.\ D} {\bf 31}, 1777 (1985)}.
  
\bibitem{Kie87} \textsc{C.~Kiefer},
  \doi{10.1088/0264-9381/4/5/031}{\jr{Class.\ Quantum Grav.} {\bf 4}, 1369 (1987)}.

\bibitem{KiS}   \textsc{C.~Kiefer} and \textsc{T.\,P.~Singh},
 \doi{10.1103/PhysRevD.44.1067}{\jr{Phys.\ Rev.\ D} {\bf 44}, 1067 (1991)}.
  
\bibitem{KK1}   \textsc{C.~Kiefer} and \textsc{M.~Kr\"amer},
  \doi{10.1103/PhysRevLett.108.021301}{\jr{Phys.\ Rev.\ Lett.} {\bf 108}, 021301 (2012)}
  [\arX{1103.4967}].
  
\bibitem{KK2}   \textsc{C.~Kiefer} and \textsc{M.~Kr\"amer}, 
  \doi{10.1142/S0218271812410015}{\jr{Int.\ J.\ Mod.\ Phys.\ D} {\bf 21}, 1241001 (2012)} [\arX{1205.5161}].

\bibitem{TMB}   \textsc{S.~Tsujikawa}, \textsc{R.~Maartens}, and \textsc{R.~Brandenberger}, 
 \doi{10.1016/j.physletb.2003.09.022}{\jr{Phys.\ Lett.\ B} {\bf 574}, 141 (2003)} [\arX{astro-ph/0308169}].

\bibitem{PTZ}   \textsc{Y.\,-S.~Piao}, \textsc{S.~Tsujikawa}, and \textsc{X.~Zhang}, 
 \doi{10.1088/0264-9381/21/18/011}{\jr{Class.\ Quantum Grav.} {\bf 21}, 4455 (2004)} [\arX{hep-th/0312139}].

\bibitem{Cal04} \textsc{G.~Calcagni},
 \doi{10.1103/PhysRevD.70.103525}{\jr{Phys.\ Rev.\ D} {\bf 70}, 103525 (2004)} [\arX{hep-th/0406006}].

\bibitem{CT1}   \textsc{G.~Calcagni} and \textsc{S.~Tsujikawa}, 
 \doi{10.1103/PhysRevD.70.103514}{\jr{Phys.\ Rev.\ D} {\bf 70}, 103514 (2004)} [\arX{astro-ph/0407543}].

\bibitem{Lar10}  \textsc{D.~Larson} {\it et al.}, \doi{10.1088/0067-0049/192/2/16}{\jr{Astrophys.\ J.\ Suppl.} {\bf 192}, 16 (2011)}
  [\arX{1001.4635}].
  
\bibitem{var1}  \textsc{J.\,R.~Bond} and \textsc{G.~Efstathiou},
\href{http://adsabs.harvard.edu/abs/1987MNRAS.226..655B}{\cob \jr{Mon.\ Not.\ Roy.\ Astron.\ Soc.} {\bf 226}, 655 (1987)}.

\bibitem{var2}  \textsc{L.~Knox} and \textsc{M.\,S.~Turner},
\doi{10.1103/PhysRevLett.73.3347}{\jr{Phys.\ Rev.\ Lett.} {\bf 73}, 3347 (1994)}
[\arX{astro-ph/9407037}].  
  
\bibitem{Kom10} \textsc{E.~Komatsu} {\it et al.}, \doi{10.1088/0067-0049/192/2/18}{\jr{Astrophys.\ J.\ Suppl.} {\bf 192}, 18 (2011)} [\arX{1001.4538}].

\bibitem{bo609} \textsc{M.\ Bojowald}, 
\doi{10.1007/s10714-006-0348-4}{\jr{Gen.\ Rel.\ Grav.} {\bf 38}, 1771 (2006)} [\arX{gr-qc/0609034}].

\bibitem{BC}    \textsc{M.~Bojowald} and \textsc{G.~Calcagni},
 \doi{10.1088/1475-7516/2011/03/032}{\jr{JCAP} {\bf 1103}, 032 (2011)} [\arX{1011.2779}].

\bibitem{BCT2}  \textsc{M.~Bojowald}, \textsc{G.~Calcagni}, and \textsc{S.~Tsujikawa},
 \doi{10.1088/1475-7516/2011/11/046}{\jr{JCAP} {\bf 1111}, 046 (2011)} [\arX{1107.1540}].

\bibitem{ALS}   \textsc{M.~Artymowski}, \textsc{Z.~Lalak}, and \textsc{\L.~Szulc},
 \doi{10.1088/1475-7516/2009/01/004}{\jr{JCAP} {\bf 0901}, 004 (2009)} [\arX{0807.0160}].

\bibitem{AKL} \textsc{A.~Ashtekar}, \textsc{W.~Kami\'nski}, and \textsc{J.~Lewandowski},
 \doi{10.1103/PhysRevD.79.064030}{\jr{Phys.\ Rev.\ D} {\bf 79}, 064030 (2009)}
  [\arX{0901.0933}].

\bibitem{WiE1} \textsc{E.~Wilson-Ewing}, 
 \doi{10.1088/0264-9381/29/8/085005}{\jr{Class.\ Quantum Grav.} {\bf 29}, 085005 (2012)} [\arX{1108.6265}].
   
\bibitem{AAN} \textsc{I.~Agull\'o}, \textsc{A.~Ashtekar}, and \textsc{W.~Nelson}, 

\arX{1204.1288}.

\bibitem{FMO} \textsc{M.~Fern\'andez-M\'endez}, \textsc{G.\,A.~Mena Marug\'an}, and \textsc{J.~Olmedo}, 
 \doi{10.1103/PhysRevD.86.024003}{\jr{Phys.\ Rev.\ D} {\bf 86}, 024003 (2012)} [\arX{1205.1917}].
 
\bibitem{WiE2} \textsc{E.~Wilson-Ewing},
 \doi{10.1088/0264-9381/29/21/215013}{\jr{Class.\ Quantum Grav.} {\bf 29}, 215013 (2012)} [\arX{1205.3370}].

\bibitem{Lan94} \textsc{D.~Langlois}, 
 \doi{10.1088/0264-9381/11/2/011}{\jr{Class.\ Quantum Grav.} {\bf 11}, 389 (1994)}.
 
\bibitem{ew1}   \textsc{M.~Bojowald}, 
 \doi{10.1103/PhysRevLett.89.261301}{\jr{Phys.\ Rev.\ Lett.} {\bf 89}, 261301 (2002)} [\arX{gr-qc/0206054}].

\bibitem{ew2}   \textsc{S.~Tsujikawa}, \textsc{P.~Singh}, and \textsc{R.~Maartens},
\doi{10.1088/0264-9381/21/24/006}{\jr{Class.\ Quantum Grav.} {\bf 21}, 5767 (2004)} [\arX{astro-ph/0311015}].

\bibitem{ew3}   \textsc{M.~Bojowald}, \textsc{J.\,E.~Lidsey}, \textsc{D.\,J.~Mulryne}, \textsc{P.~Singh}, and \textsc{R.~Tavakol},
 \doi{10.1103/PhysRevD.70.043530}{\jr{Phys.\ Rev.\ D} {\bf 70}, 043530 (2004)} [\arX{gr-qc/0403106}].

\bibitem{ew4}   \textsc{G.\,M.~Hossain},
 \doi{10.1088/0264-9381/22/12/012}{\jr{Class.\ Quantum Grav.} {\bf 22}, 2511 (2005)} [\arX{gr-qc/0411012}].

\bibitem{ew5}   \textsc{G.~Calcagni} and \textsc{M.~Cort\^es},
 \doi{10.1088/0264-9381/24/4/005}{\jr{Class.\ Quantum Grav.} {\bf 24}, 829 (2007)} [\arX{gr-qc/0607059}].

\bibitem{ew6}   \textsc{E.\,J.~Copeland}, \textsc{D.\,J.~Mulryne}, \textsc{N.\,J.~Nunes}, and \textsc{M.~Shaeri},
 \doi{10.1103/PhysRevD.77.023510}{\jr{Phys.\ Rev.\ D} {\bf 77}, 023510 (2008)} 

 [\arX{0708.1261}].

\bibitem{ew7}   \textsc{M.~Shimano} and \textsc{T.~Harada},
 \doi{10.1103/PhysRevD.80.063538}{\jr{Phys.\ Rev.\ D} {\bf 80}, 063538 (2009)} [\arX{0909.0334}].

\bibitem{BH1}   \textsc{M.~Bojowald} and \textsc{G.\,M.~Hossain}, 
 \doi{10.1088/0264-9381/24/18/015}{\jr{Class.\ Quantum Grav.} {\bf 24}, 4801 (2007)} [\arX{0709.0872}].

\bibitem{BH2}   \textsc{M.~Bojowald} and \textsc{G.\,M.~Hossain}, 
 \doi{10.1103/PhysRevD.77.023508}{\jr{Phys.\ Rev.\ D} {\bf 77}, 023508 (2008)} [\arX{0709.2365}].

\bibitem{BHKS}  \textsc{M.~Bojowald}, \textsc{G.\,M.~Hossain}, \textsc{M.~Kagan}, and \textsc{S.~Shankaranarayanan}, 
 \doi{10.1103/PhysRevD.78.063547}{\jr{Phys.\ Rev.\ D} {\bf 78}, 063547 (2008)} [\arX{0806.3929}].

\bibitem{BHKS2} \textsc{M.~Bojowald}, \textsc{G.M.~Hossain}, \textsc{M.~Kagan}, and \textsc{S.~Shankaranarayanan}, 
 \doi{10.1103/PhysRevD.79.043505}{\jr{Phys.\ Rev.\ D} {\bf 79}, 043505 (2009)} [\doi{10.1103/PhysRevD.82.109903}{\jr{Erratum ibid.\ D} {\bf 82}, 109903(E) (2010)}] [\arX{0811.1572}].

\bibitem{CMNS2} \textsc{E.\,J.~Copeland}, \textsc{D.\,J.~Mulryne}, \textsc{N.\,J.~Nunes}, and \textsc{M.~Shaeri}, 
 \doi{10.1103/PhysRevD.79.023508}{\jr{Phys.\ Rev.\ D} {\bf 79}, 023508 (2009)}
 
  [\arX{0810.0104}].

\bibitem{CH}    \textsc{G.~Calcagni} and \textsc{G.\,M.~Hossain}, 
 \doi{10.1166/asl.2009.1025}{\jr{Adv.\ Sci.\ Lett.} {\bf 2}, 184 (2009)} [\arX{0810.4330}].

\bibitem{BCT1}  \textsc{M.~Bojowald}, \textsc{G.~Calcagni}, and \textsc{S.~Tsujikawa}, 
 \doi{10.1103/PhysRevLett.107.211302}{\jr{Phys.\ Rev.\ Lett.} {\bf 107}, 211302 (2011)} [\arX{1101.5391}].

\bibitem{Cal12} \textsc{G.~Calcagni},
 \doi{10.1088/1742-6596/360/1/012027}{\jr{J.\ Phys.\ Conf.\ Ser.} {\bf 360}, 012027 (2012)}
 [\arX{1110.0291}].
 
\bibitem{MFB}   \textsc{V.\,F.~Mukhanov}, \textsc{H.\,A.~Feldman}, and \textsc{R.\,H.~Brandenberger},  
  \doi{10.1016/0370-1573(92)90044-Z}{\jr{Phys.\ Rept.} {\bf 215}, 203 (1992)}.

\bibitem{Antony} \url{http://cosmologist.info/cosmomc/}

\bibitem{Reid} \textsc{B.\,A.~Reid} {\it et al.}, 
 \doi{10.1111/j.1365-2966.2010.16276.x}{\jr{Mon.\ Not.\ Roy.\ Astron.\ Soc.} \textbf{404}, 60 (2010)} [\arX{0907.1659}].

\bibitem{HST} \textsc{A.\,G.~Riess} {\it et al.}, 
\doi{10.1088/0004-637X/699/1/539}{\jr{Astrophys.\ J.} \textbf{699}, 539 (2009)} [\arX{0905.0695}].

\bibitem{SNIa} \textsc{M.~Kowalski} {\it et al.} {[}Supernova Cosmology
Project Collaboration{]}, 
\doi{ 	10.1086/589937}{\jr{Astrophys.\ J.} \textbf{686}, 749 (2008)} [\arX{0804.4142}].

\bibitem{BBN} \textsc{S.~Burles} and \textsc{D.~Tytler}, 
\doi{10.1086/305667}{\jr{Astrophys.\ J.} \textbf{499}, 699 (1998)} [\arX{astro-ph/9712108}].

\bibitem{APSII}  \textsc{A.~Ashtekar}, \textsc{T.~Pawlowski}, and \textsc{P.~Singh},
 \doi{10.1103/PhysRevD.74.084003}{\jr{Phys.\ Rev.\ D} {\bf 74}, 084003 (2006)} 
[\arX{gr-qc/0607039}].

\bibitem{Boj74} \textsc{M.~Bojowald},
 \doi{10.1103/PhysRevD.75.081301}{\jr{Phys.\ Rev.\ D} {\bf 74}, 081301 (2007)} [\arX{gr-qc/0608100}].

\bibitem{BNMR}  \textsc{M.~Bojowald}, \textsc{W.~Nelson}, \textsc{D.~Mulryne}, and \textsc{R.~Tavakol},
 \doi{10.1103/PhysRevD.82.124055}{\jr{Phys.\ Rev.\ D} {\bf 82}, 124055 (2010)} [\arX{1004.3979}].

\bibitem{MCBG}  \textsc{J.~Mielczarek}, \textsc{T.~Cailleteau}, \textsc{A.~Barrau}, and \textsc{J.~Grain}, 

 \doi{10.1088/0264-9381/29/8/085009}{\jr{Class.\ Quantum Grav.} {\bf 29}, 085009 (2012)} 
   
 [\arX{1106.3744}].

\bibitem{CMBG}  \textsc{T.~Cailleteau}, \textsc{J.~Mielczarek}, \textsc{A.~Barrau}, and \textsc{J.~Grain}, 
 \doi{10.1088/0264-9381/29/9/095010}{\jr{Class.\ Quantum Grav.} {\bf 29}, 095010 (2012)} [\arX{1111.3535}].

\bibitem{CaB}   \textsc{T.~Cailleteau} and \textsc{A.~Barrau},
  \doi{10.1103/PhysRevD.85.123534}{\jr{Phys.\ Rev.\ D} {\bf 85}, 123534 (2012)} [\arX{1111.7192}].
  
\bibitem{CBGV}  \textsc{T.~Cailleteau}, \textsc{A.~Barrau}, \textsc{J.~Grain}, and \textsc{F.~Vidotto},
  \doi{10.1103/PhysRevD.86.087301}{\jr{Phys.\ Rev.\ D} {\bf 86}, 087301 (2012)} [\arX{1206.6736}].

\bibitem{BoP}  \textsc{M.~Bojowald} and \textsc{G.\,M.~Paily},  
 \doi{10.1088/0264-9381/29/24/242002}{\jr{Class.\ Quantum Grav.} {\bf 29}, 242002 (2012)} [\arX{1206.5765}]. 

\bibitem{Mie1}  \textsc{J.~Mielczarek}, 
 \doi{10.1088/1475-7516/2008/11/011}{JCAP {\bf 0811}, 011 (2008)} 
 
 [\arX{0807.0712}].

\bibitem{GB1}   \textsc{J.~Grain} and \textsc{A.~Barrau}, 
 \doi{10.1103/PhysRevLett.102.081301}{\jr{Phys.\ Rev.\ Lett.} {\bf 102}, 081301 (2009)} [\arX{0902.0145}].

\bibitem{GB2}   \textsc{J.~Mielczarek}, \textsc{T.~Cailleteau}, \textsc{J.~Grain}, and \textsc{A.~Barrau},
 \doi{10.1103/PhysRevD.81.104049}{\jr{Phys.\ Rev.\ D} {\bf 81}, 104049 (2010)} [\arX{1003.4660}].

\bibitem{GB3}   \textsc{J.~Grain}, \textsc{A.~Barrau}, \textsc{T.~Cailleteau}, and \textsc{J.~Mielczarek},
 \doi{10.1103/PhysRevD.82.123520}{\jr{Phys.\ Rev.\ D} {\bf 82}, 123520 (2010)} [\arX{1011.1811}].  

\bibitem{LCGH} \textsc{L.-F.~Li}, \textsc{R.-G.~Cai}, \textsc{Z.-K.~Guo}, and \textsc{B.~Hu}, \doi{10.1103/PhysRevD.86.044020}{\jr{Phys.\ Rev.\ D} {\bf 86}, 044020 (2012)} [\arX{1112.2785}].

\bibitem{GLMM}  \textsc{A.~Gangui}, \textsc{F.~Lucchin}, \textsc{S.~Matarrese}, and \textsc{S.~Mollerach},
  \doi{10.1086/174421}{\jr{Astrophys.\ J.} {\bf 430}, 447 (1994)} [\arX{astro-ph/9312033}].
  
\bibitem{VWHK}  \textsc{L.~Verde}, \textsc{L.-M.~Wang}, \textsc{A.~Heavens}, and \textsc{M.~Kamionkowski},
  \doi{10.1046/j.1365-8711.2000.03191.x}{\jr{Mon.\ Not.\ Roy.\ Astron.\ Soc.} {\bf 313}, L141 (2000)} [\arX{astro-ph/9906301}].
  
\bibitem{KoS}   \textsc{E.~Komatsu} and \textsc{D.\,N.~Spergel},
  \doi{10.1103/PhysRevD.63.063002}{\jr{Phys.\ Rev.\ D} {\bf 63}, 063002 (2001)} [\arX{astro-ph/0005036}].

\bibitem{Mal02} \textsc{J.\,M.~Maldacena},
  \doi{10.1088/1126-6708/2003/05/013}{\jr{JHEP} {\bf 0305}, 013 (2003)} 
  
  [\arX{astro-ph/0210603}].
  
\bibitem{BCZ}   \textsc{D.~Babich}, \textsc{P.~Creminelli}, and \textsc{M.~Zaldarriaga},
  \doi{10.1088/1475-7516/2004/08/009}{\jr{JCAP} {\bf 0408}, 009 (2004)} [\arX{astro-ph/0405356}].
  
\bibitem{CrZ}   \textsc{P.~Creminelli} and \textsc{M.~Zaldarriaga}, 
  \doi{10.1088/1475-7516/2004/10/006}{\jr{JCAP} {\bf 0410}, 006 (2004)} [\arX{astro-ph/0407059}].
  
\bibitem{SHSHP} \textsc{A.~Slosar}, \textsc{C.~Hirata}, \textsc{U.~Seljak}, \textsc{S.~Ho}, and \textsc{N.~Padmanabhan},
  \doi{10.1088/1475-7516/2008/08/031}{\jr{JCAP} {\bf 0808}, 031 (2008)} [\arX{0805.3580}].
 }
\end{thebibliography}

\end{document}